\documentclass[aps,prd,showpacs,nofootinbib,preprintnumbers,twocolumn]{revtex4-1}
\usepackage[ansinew]{inputenc} 
\usepackage{graphicx,epsfig}
\usepackage{url}
\usepackage{color}
\usepackage[colorlinks=true,linkcolor=black,allcolors=red]{hyperref}
\hypersetup{colorlinks,linkcolor={blue},citecolor={red},urlcolor={red}}
\usepackage{bm}
\usepackage{dcolumn}
\usepackage{amsfonts,amssymb,amsmath}
\usepackage{psfrag}
\usepackage{subfigure}
\usepackage{natbib}
\usepackage{latexsym}
\begin{document}
\title{Constraining $f(R)$ model through spectral indices and reheating temperature}
\author{Ajay Sharma}
\email[]{aksh.sharma2@gmail.com}
\affiliation{Department of Physics, University of Lucknow, Lucknow, 226007, India}
\author{Murli Manohar Verma}
\email[]{sunilmmv@yahoo.com}
\affiliation{Department of Physics, University of Lucknow, Lucknow, 226007, India}
\date{\today}
\begin{abstract}
 We investigate a   form of $ f(R) = {R^{1+\delta}}/{R_c^{\delta}}$  and study  the viability of the model for inflation in the Jordan and the Einstein frames. This model is further analysed  by using the power spectrum indices of the inflation and the reheating temperature. During the inflationary evolution, the model predicts a value of $\delta$  parameter  very close to one  ($\delta=0.98$),  while the reheating temperature  $T_{re} \sim 10^{17}$ GeV at $\delta=0.98$  is  consistent with the standard approach to inflation and observations.  We  calculate the slow roll parameters for the  minimally coupled scalar field within  the framework of  our  model.  It is found that the   values  of the scalar spectral index and tensor-to-scalar ratio  are very close to the recent observational data, including those  released by Planck 2018.  We also show  that the Jordan and the Einstein frames are  equivalent  when $\delta \sim 1 $  by using  the scalar spectral index, tensor-to-scalar ratio and reheating temperature.
\end{abstract}
\maketitle
\section{\label{1}Introduction}
In study of  the very early universe,  inflation was introduced to solve the horizon problem, flatness problem, monopole problem, entropy problem etc.
These are among the most pronounced problems of the Lambda Cold Dark Matter ($\Lambda$CDM)  model in  research  in cosmology at present. Of course,  even though there  exist  several competent  solutions of  these problems of $\Lambda$CDM model, still  we do not have a  completely viable inflationary model. In  literature, there are several  models like the  Starobinsky model, Chaotic inflationary model, Plateau type inflationary model etc.  which attempt to solve  these  issues. Among these models,  the  Starobinsky model has its own merits to be considered as  the   most significant  one \cite{I1}.

Several inflationary models  use  the scalar field having  constant energy density during inflation just like the cosmological constant or the vacuum energy density \cite{I2}. Models  of    cosmological constant varying   through    interaction with the background in an  intermediate phase sandwiched  between the early inflation and the present accelerated phase  have also been proposed \cite{I2111}.  Some  authors assume that the universe was supercooled as vacuum in the very early universe. Its source was  considered to be the entropy \cite{I3}. Following this,    Guth proposed  an  inflationary model in 1981 \cite{I4}.  It was also based on  supercooling in the false vacuum state  where  the universe enters  a  reheating phase by means of  bubble collision \cite{I5}.  However,  this approach does not work well because it is inflicted from  the reheating problem which  needs to be solved.



A viable inflationary model should be able to reheat the universe. Reheating begins   after the end of inflationary phase and  this  phase is very crucial for our universe because it  increases the temperature of the very cold universe.  The   Grand Unification Theory (GUT) energy scale lies  between $10^{13}$ GeV to $10^{16}$ GeV and electroweak Spontaneous Symmetry  Breaking (SSB) phase transition occurs at $\sim 300$ GeV, therefore reheating temperature should be  $> 10^{16}$ GeV \cite{I5.1, I5.2}. According to the particle field theories and nuclear synthesis, temperature of the universe should be greater than $100$ GeV  after the reheating process ends \cite{I5.2}.  Till 1982, there was a lapse of the viable model which  could  solve the problems of the $\Lambda$CDM model as well as graceful exit problem of the old inflationary model.

 In 1982,  Linde proposed  a inflationary model \cite{I6},  known as `new' inflationary  model. This model offers   the solution to  the $\Lambda$CMD model's problems,  including the  graceful exit problem. The basic difference between old and new inflationary models is that the universe becomes homogeneous in the new inflationary model whereas it was inhomogeneous in the old inflationary theory.  The recent observations of  Cosmic Microwave Background power spectrum is uniform at the order of $10^{-5}$ \cite{I7,I7.1},  so that the new inflationary model is more successful than the old inflationary model.   Linde proposed  a chaotic inflationary theory also, in 1983 \cite{I8}.

There is a class  of  theories to explain the inflation based upon the scalar field theory and  modified gravity theories \cite{I1,I9,I10,I11,I12,I13}.  Starobinsky's inflationary model became a  viable model, particularly   after release of the  Planck 2018 data \cite{I14}.   Starobinsky proposed the  first modified gravity model for inflation in 1980 \cite{I1} and several  other authors also published  valuable work  on  inflation in  the framework of  modified gravity models \cite{I15, I16, I17,I17.1, I18}. Scalar-tensor  theory is a modified gravity model,  and  for the first time,  Brans and  Dicke introduced a scalar-tensor theory by replacing inverse of the Newtonian gravitational constant $G^{-1}$ by a  scalar field $\phi$ \cite{I18}.

We transform the spacetime metric $g_{\mu\nu}$ from the Jordan frame to the Einstein frame by the conformal transformation. In the scalar-tensor theory, the scalar field arises as a new degree of freedom after conformal transformation of the metric tensor $g_{\mu\nu}$.  In  cosmology,  there is a serious debate on the equivalence of the  Jordan and the Einstein frames.  Several  approaches have been used to understand  and settle down this  problem \cite{I18.1,I18.2,I18.3,I18.4,I18.5}.

In the present  paper, we consider a general function of Ricci scalar, $R$, given by
\begin{eqnarray}f(R) = \frac{R^{1+\delta}}{R_c^\delta}, \label{I1} \end{eqnarray}
where $R_c$ is a constant  and   $\delta$ is a dimensionless model parameter.  This  $f(R)$ gravity model has  also been used to explain the dark matter problem in \cite{I20}.  If $\delta\rightarrow 0$, we   recover the Einstein-Hilbert action as  $ f(R) = {R^{1+\delta}}/{R_c^{\delta}} \rightarrow R$.   This form of $f(R)$ is inspired by the model $f(R) = R^{1+\delta}$ \cite{I21},  where the parameter $\delta$  is a function of the constant tangential velocity  and its value,  as  calculated from the constant tangential velocity of the test particles ($200-300$ km/s),    is  of the order of $(v^2/c^2)\sim10^{-6}$.  The  viability of a similar  f(R) model with a small  $\delta$  has also  been discussed  for extended  galactic rotational velocity profiles in weak field  approximation \cite{I211}.    Therefore, a small correction in $R$  may explain the issues which are otherwise  solved by invoking  dark matter.  However, as we will notice  further  in this paper,   we require a  larger value of $\delta$  compared   to $10^{-6}$  to account for  inflation at the early epoch in the universe.  We  use  the $f(R)$ model given by equation \eqref{I1} to address the issues of inflation and reheating in modified gravity scanario .

The inflationary observational parameters  as  the  scalar spectral index $n_s$, tensor-to-scalar ratio $r$,  amplitude of the scalar power spectrum $A_s$ etc.  are constraints on model's parameters and the reheating temperature \cite{I14}. These are calculated from the Cosmic Microwave Background (CMB) power spectrum of the universe, mainly  through  the   Planck's observations, Wilkinson  Microwave Anisotropy  Probe (WMAP) observations etc.

The present  paper is organised as follows. In section II, we  focus our attention to   inflation, reheating temperature and power spectral indices in the scalar-tensor theory.  We  also calculate the  value of the $\delta$ parameter in both frame the Jordan and  the Einstein frame in sections  III and  IV,  respectively.  We  compare  the Jordan and the Einstein frames in the section V and discuss the  conclusion in section VI.

  Further, we  use  the  Greek letters $\mu,\nu,\alpha,\beta = 0, 1, 2, 3 $ and Latin letters $i,j,k = 1, 2, 3$. The sign convention used  for metric is $(-, + , + , + )$ and the  natural unit system $c =\hbar = k_B = 1 $, where c, $\hbar$ and $k_B$  are  the speed of light, reduced Planck constant and the  Boltzmann's constant, respectively. Dot and prime are considered   as  differentiation w.r.t. cosmic time  and Ricci scalar, respectively.

\section{\label{1.1}Inflationary dynamics and Reheating temperature }
In the present work, we  consider the scalar-tensor theory as an effective  theory and study  inflation within it.
We  further assume  the background  dynamic universe  as being   homogeneous, isotropic  and   spatially flat, which is defined by the Friedmann-Robertson-Walker (FRW) spacetime metric,   given by the spacetime interval,
\begin{eqnarray}ds^{2}= -dt^{2} + a^2(t)[dr^2 + r^2 (d\theta^{2} + \sin^{2}\theta d\phi^{2})], \label{IA1} \end{eqnarray}
where  $a(t)$ is scale factor, $t$ is the cosmic time and $r, \theta,\phi$ are the spherical coordinates.

 We know that the strong energy condition is $\rho + 3P \geq 0$,  so that  the universe undergoes  deceleration. The Equation of State (EoS) $w$ of  all the known normal matter components of the universe is $ \geq -1/3 $.  However,  an  interpretation  of the  observational data of Supernovae type Ia (SN-Ia) in 1998   suggests    that the present universe is in an accelerating phase \cite{Ia1,Ia2}.

 From the above discussion, we can conclude that the normal matter does not produce accelerated expansion of the universe. However,  the acceleration  can  still  be obtained   by adding some exotic matter   or making  `correction'  in the curvature part of the Einstein-Hilbert action. There are some other ways  to solve this problem  in  emergent gravity models, steady state theory, string theory etc. The aim of this `correction' is to attain  the accelerated expansion of  the universe  with    the EoS of the matter  $ w < -(1/3) $. Such an  equation of state can be  obtained by the violation of the strong energy condition  i.e., by  ensuring   $ \rho + 3P < 0$. In this paper,   we   discuss  about the inflation in a  de-Sitter spacetime.

 Such a universe in the present  accelerated phase   can be realised  by  the  $\Lambda$ cosmological constant. Of course,  the $\Lambda$  model cannot be a cosmologically viable model for the accelerated expansion in the early universe because exponential expansion will never end \cite{Ia3}.  Therefore, we do not need exactly constant energy density  in the early universe.  Actually,   we need approximately constant energy density i.e $ H\simeq $ constant  and quasi de-Sitter spacetime during inflation.

We can write  ${\ddot a}/{a} $ in terms of $H$ and it derivatives
 \begin{eqnarray} \dfrac{\ddot a}{a}= H^2 + \dot H = H^2\Big(1 + \dfrac{\dot H}{H^2}\Big), \label{IA7}\end{eqnarray}
where $H\equiv\dot a/a$ is the Hubble parameter and $\dot H$ is the time derivative of the Hubble parameter.
 Now, we define
 \begin{eqnarray} \varepsilon_1\equiv -\dfrac{\dot H}{H^2}. \label{IA8}\end{eqnarray}
 We  get acceleration from equations \eqref{IA7} and \eqref{IA8}
\begin{align}
H^2(1 - \varepsilon_1) > 0  ;
&& if
&& \varepsilon_1 < 1. \label{IA9}\end{align}
It means that the acceleration can be produced if and only if $\varepsilon_1 < 1 $ i.e. $|\dot H/H^2|<1$.

Here,  we have defined the parameters  for slow roll inflation \cite{Ia4,Ia5}
\begin{align}
\varepsilon_1\equiv\frac{-\dot H}{H^2} ;
&&  \eta_H\equiv\frac{-\ddot H}{\dot H H} ;
\label{IA24}\end{align}
 and these parameters  should be
\begin{align}
\varepsilon_1\ll 1 ;
&&  \eta_H\ll 1.
\label{IA24.1}\end{align}
To explain inflation in the scalar field description, we  assume  a homogenous self-interacting scalar field $\phi(t)$ having potential $ V(\phi)$, minimally coupled to  the curvature in the very early universe \cite{Ia6,Ia7}.   Then,  the slow roll parameters can be  defined as
\begin{align}
\varepsilon_1\equiv\frac{\dot\phi^2}{2V(\phi)};
&&  \varepsilon_2\equiv\frac{\ddot \phi}{H\dot\phi }.
\label{IA26}\end{align}
To study the inflation,  using the scalar field potential is quite  convenient. Therefore,  we have the slow roll parameters $\varepsilon_1$ and $\varepsilon_2$ in terms of $V(\phi)$,
\begin{eqnarray}
\varepsilon_1 \equiv \dfrac{M^2_{pl}}{2}\Big(\dfrac{V,_\phi}{V}\Big)^2; \label{IA32}\end{eqnarray}
\begin{eqnarray}\varepsilon_2  \equiv \dfrac{M^2_{pl}}{2}\Big(\dfrac{V,_\phi}{V}\Big)^2- M^2_{pl}\Big(\dfrac{V,_{\phi\phi}}{V}\Big),  \label{IA33}\end{eqnarray}
and  we  also  define another parameter
\begin{eqnarray}\eta\equiv  M^2_{pl}\Big(\dfrac{V,_{\phi\phi}}{V}\Big). \label{IA34}\end{eqnarray}
These parameters for inflation should be
\begin{align}
\varepsilon_1\ll 1;
&& \varepsilon_2\ll 1;
&&  \eta \ll 1.
\label{IA33.1}\end{align}
Now,  using the equations \eqref{IA32}, \eqref{IA33} and \eqref{IA34}, we obtain
\begin{eqnarray} \varepsilon_2 = \varepsilon_1 - \eta. \label{IA35} \end{eqnarray}
We have the number of e-folds $N = \ln[a(t_{en})/a(t)]$,  where $t_{en}$ is the epoch   at which the inflation ends \cite{Ia6}, given as
\begin{eqnarray} N =  \dfrac{1}{M_p}\int^{t_{f}}_{t_i} H dt = \dfrac{1}{M_p}\int^{\phi_{end}}_{\phi_{i}}{\dfrac{H}{\dot\phi} d\phi}. \label{IA36}\end{eqnarray}
We assume    that the  number of e-foldings required  to solve the problems of the $\Lambda$CDM model  are $50<  N <60$ \cite{Ia8,I14}. Therefore,
\begin{eqnarray} N = \int^{\phi_{60}}_{\phi_{end}}{\dfrac{d(\phi/M_p)}{\sqrt{2\varepsilon_1}} }. \label{IA37}\end{eqnarray}
The power spectrum indices, scalar spectral index $n_s$ and tensor-to-scalar ratio $r$ are given as \cite{Ia5,Ia9}
\begin{eqnarray}  n_s = 1- 6\varepsilon_1 + 2\eta \label{IA38} \end{eqnarray}
and
\begin{eqnarray}   r = 16 \varepsilon_1.  \label{IA39}\end{eqnarray}
Equations \eqref{IA37}, \eqref{IA38} and \eqref{IA39} are used to calculate the values of $ N$, $n_s$ and $r$,  respectively,  in the present   model \eqref{I1}. These are the significant  parameters of the inflation.

Observational values of these parameters from the recent Planck 2018 data analysis\cite{I14} are
\begin{align} n_{s}  = 0.9649 \pm 0.0042,
&& at && 68\% && CL; \label{IA40} \end{align}
\begin{align} r < 0.1,
&& at && 95\% && CL; \label{IA41}  \end{align}
and by BICEP2/Keck Array BK14 recent data\cite{Ia10}
\begin{align} r < 0.065,
&& at &&  95\% &&CL. \label{IA42}\end{align}
In this paper,  we   compare the calculated values of the power spectrum indices  with  the observations  and check the consistency of this closeness.

\subsection{\label{1.3} Reheating temperature $T_{re}$}

A viable inflationary models should be able to reheat the universe. Reheating starts after the end of inflationary phase. This phase is quite  crucial for our universe because it  increases the temperature of the super cold universe, which is necessary for further evolution.  Therefore,   we now  calculate the reheating temperature.

We assume  that the entropy   ($S = sV$, where $S$   is  entropy, $s$   is entropy density  and $V$  is volume of the patch) does not change during adiabatic expansion after the  end of the reheating till now,  i.e.,  $s_{re}a_{re}^3= s_o a_o^3$,
where $s_{re} $, $s_o$, $a_{re}$  and  $a_o $  denote  entropy density at the end of reheating, today's entropy density, scale factor at the end of reheating and  present  scale factor, respectively. We can thus  connect  the reheating and present epochs   via entropy.

After reheating, universe was  dominated by the relativistic species,  and  the   total entropy   was almost carried by the these relativistic species till the present epoch.  Entropy density of the relativistic species can be given as  \cite{I10}
\begin{eqnarray}s_{re} = \frac{2\pi^2}{45}g_{s,re}T_{re}^3, \label{IC3} \end{eqnarray}
where $g_{s,re}$  is  total relativistic degree of  freedom at end of reheating and $T_{re}$ is the reheating temperature. Today,  the entropy density of  relativistic species is
\begin{eqnarray}s_o = \frac{2\pi^2}{45}g_{s,o}T_{o}^3, \label{IC4}\end{eqnarray}
where $g_{s,o} = (43/11)$ \cite{I10} refers to the total number of relativistic degrees  of freedom after neutrino decoupling and $T_{o}$ is CMB temperature today. Now,   using the relation \eqref{IC3} and \eqref{IC4}, we have
\begin{eqnarray}T_{re} = \Big(\frac{43}{11g_{s,re}}\Big)^{1/3} \frac{a_o T_o}{a_{re}} , \label{IC5}\end{eqnarray}
All the perturbations get frozen when wavelengths  exit the Hubble horizon at the wave number $k = a_k H_k$, where $a_k$ and $H_k$ are the scale factor and Hubble parameter,  respectively,  at the exit of the Hubble horizon. These perturbations are  imprinted on the Cosmic Microwave Background (CMB)  and their  signatures are observed  when wavelength re-enter the Hubble horizon at the wave number $k = a_o H_o$, where $a_o$ and $H_o$ are the scale factor and Hubble parameter,  respectively,  at the re-entry  of the Hubble horizon.
We have the relation to calculate the value of $ a_o/a_{re}$
\begin{eqnarray}\dfrac{k}{a_k H_k} = \frac{a_{en}}{a_k} \frac{a_{re}}{a_{en}} \frac{a_o}{a_{re}} \frac{k}{a_o H_k},  \label{IC6}\end{eqnarray}
where $a_{en}$ is the scale factor at the end of inflation.  Using the definition of the number of e-foldings \cite{Ia6} for $N_k \equiv \ln{({a_{en}}/{a_{k}})}$ and $ N_{re} \equiv \ln{({a_{re}}/{a_{en}})}$. We have
\begin{eqnarray} \frac{a_o}{a_{re}} = e^{-N_k-N_{re}} \frac{a_o H_k}{k},  \label{IC7}\end{eqnarray}
 and putting the value of ${a_o}/{a_{re}}$ in the equation \eqref{IC5} from equation \eqref{IC7}
\begin{eqnarray}T_{re} = \Big(\frac{43}{11g_{s,re}}\Big)^{1/3} \frac{a_o T_o}{k} H_k e^{-N_k-N_{re}}. \label{IC8}\end{eqnarray}

As usual energy density of the universe vary $\rho_i \propto a^{-3(1+w_i)}$, here $w_i$ is a EoS of the fluid and $i$ refers to different species. Therefore we have a relation between  $\rho_{en}$ and $\rho_{re}$ as
\begin{eqnarray}\dfrac{\rho_{re}}{\rho_{en}} = \Big(\dfrac{a_{re}}{a_{en}}\Big)^{-3(1+w_{re})}, \label{IC9}\end{eqnarray}
where $ \rho_{en}$ is a energy density at the end of inflation, $ \rho_{re}$ is a energy density at the end of reheating, $w_{re}$ is an equation of state during reheating.

We have a relation for reheating  number of e-foldings $N_{re}$ in terms of $ \rho_{en}$ and $ \rho_{re}$ by using the above equation \eqref{IC9}
\begin{eqnarray} N_{re} = \frac{1}{3(1+w_{re})} \ln{\Big(\frac{\rho_{en}}{\rho_{re}} \Big)}.   \label{IC10}\end{eqnarray}
Here, we assume that the total energy density of the inflaton field completely converted into energy density of the relativistic species during reheating. Therefore the energy density of the relativistic species at the end of reheating is given by
\begin{eqnarray}\rho_{re} = \dfrac{\pi^2}{30}g_{re} T_{re}^4, \label{IC11}\end{eqnarray}
where,  $g_{re}$ is  the relativistic degree of freedom.

Now,  putting the value of $\rho_{re}$ in the equation \eqref{IC10} from equation \eqref{IC11},
\begin{eqnarray}e^{-N_{re}} = \Big(\dfrac{30\rho_{en}}{\pi^2 g_{re}} \Big)^{-\frac{1}{3(1+w_{re})}} T_{re}^{\frac{4}{3(1+w_{re})}}. \label{IC13}\end{eqnarray}

and with  the value of the $e^{-N_{re}}$ from \eqref{IC13} in the equation \eqref{IC8},

\begin{eqnarray}T_{re} = \Big[\Big(\frac{43}{11g_{s,re}}\Big)^{1/3} \Big(\frac{a_o T_o}{k}\Big) H_k e^{-N_k}\Big(\dfrac{30\rho_{en}}{\pi^2 g_{re}} \Big)^{-\frac{1}{3(1+w_{re})}}\Big]^{\frac{3(1+w_{re})}{3w_{re}-1}}. \label{IC14}\end{eqnarray}
Considering   $w_{re} = 0$ during reheating,   $T_{re}$ becomes
\begin{eqnarray}T_{re} = \Big(\frac{11g_{s,re}}{43}\Big) \Big(\frac{k}{a_o T_o}\Big)^3 \Big(\dfrac{30}{\pi^2 g_{re}} \Big) H_k^{-3} e^{3N_k} \rho_{en}. \label{IC15}\end{eqnarray}
From the above equation it is clear that the reheating temperature depends on the $H_k$, $N_k$ and $\rho_{en}$. This leads us to   calculate  the reheating temperature by using these  values  in the Jordan  and  the Einstein frames.

\subsection{\label{1.2} Inflationary dynamics in the Scalar-Tensor (S-T) theory }
In the generalized scalar-tensor theories, action  without matter field can be written as \cite{Ib1}

\begin{eqnarray} \mathcal{A} =\int{ d^4 x {\sqrt{-g}}\left[ \dfrac{f(R, \phi)}{2\kappa^2} -\dfrac{\omega}{2} g^{\mu\nu}\partial_{\mu}\phi\partial_{\nu}\phi - V(\phi) \right]} \label{IB1}\end{eqnarray}
where $f(R, \phi)$ is a function of Ricci scalar and  scalar field and $\omega$ is a parameter which is $\neq 1$ for non-canonical scalar field, and  $=1$ for canonical scalar field.
Varying the action \eqref{IB1} w.r.t. the scalar field $\phi$ and metric tensor  with $\omega = 1$. We obtain  the equations of motion \cite{Ib1}

\begin{eqnarray} H^2 = \dfrac{1}{3F}\left\{\dfrac{\dot\phi^2}{2} + V(\phi) + \dfrac{FR -f }{2} - 3H\dot F \right\} \label{IB2}\end{eqnarray}
\begin{eqnarray} \dot H = -\dfrac{1}{2F}(\dot\phi^2 + \ddot F - H\dot F ) \label{IB3}\end{eqnarray}
\begin{eqnarray} \ddot\phi + 3H\dot\phi + \dfrac{1}{2}(2V_{,\phi} - f_{,\phi}) = 0, \label{IB4}\end{eqnarray}
where $V_{,\phi}= {\partial V}/{\partial\phi}$, $ f_{,\phi} = {\partial f}/{\partial\phi}$, $ F = {\partial f}/{\partial R}$.

In $f(R,\phi)$ theories, the  slow roll parameters are defined   as\cite{Ia5,Ib1}
\begin{align}\varepsilon_1 \equiv -\dfrac{\dot H}{H^2};
&& \varepsilon_2 \equiv \dfrac{\ddot \phi}{\dot\phi H};
&& \varepsilon_3 \equiv \dfrac{\dot F}{2HF};
&& \varepsilon_4 \equiv \dfrac{\dot E}{2HE}, \label{IB5}\end{align}
where
\begin{eqnarray} E \equiv F\left[1 + \dfrac{3\dot F^2}{2\kappa^2\dot\phi^2 F }\right].\label{IB6}\end{eqnarray}
We can write  slow roll parameters in the generalised form as $\varepsilon_i$, where $i = 1,2,3,4$ and we consider  $\dot\varepsilon_i \simeq 0$ \cite{Ia5}.

Cosmological perturbations in $ f(R)$ theory has been studied in \cite{Ib1}. Perturbations generated during inflation are strong evidence of the inflation. These perturbations  freeze after crossing the Hubble radius. They are imprinted on the Cosmic Microwave Background (CMB) and  used  as a fingerprint of the inflation. These fingerprints are known as power spectral indices (e.g. tensor-to-scalar ratio, scalar spectral index, tensor spectral index etc.). Power spectrum of the CMB is almost scale invariant $(n_s\sim 1)$ during inflation.\\
Spectral index $n_s$ during inflation in modified gravity theories is \cite{Ia9}
\begin{eqnarray} {n}_{s}\simeq 1-4{\varepsilon}_1 -2{\varepsilon}_2 + 2{\varepsilon}_3 - 2{\varepsilon}_4, \label{IB8}\end{eqnarray}

and tensor-to-scalar ratio $r$ is \cite{Ia9}
\begin{eqnarray} r = \dfrac{{\mathcal{P}_T}}{{\mathcal{P}_s}} \simeq \dfrac{8 \kappa^2 Q_s}{F}, \label{IB9}\end{eqnarray}

where $ Q_s = \dot\phi^2 E/[FH^2(1+\varepsilon_3)^2] $; ${\mathcal{P}_s}\simeq {Q_s}^{-1}( { H}/{2\pi} )^2 $ and ${\mathcal{P}_T}\simeq {16}( H \kappa)^2 /{\pi}{F} $ \cite{Ia9}.

Scalar curvature perturbation,
$ \mathcal{R} = \psi - {H \delta F}/{\dot F} $,  remains invariant $ \mathcal{R} = \tilde{\mathcal{R}}$ under the conformal transformation and
tensor perturbation is also remains invariant. Therefore, tensor-to-scalar ratio and scalar spectral index in the Jordan frame are  identical with Einstein frame \cite{Ia9}.

\section{\label{2} Inflationary dynamics and Reheating temperature in the Jordan frame }
Action in  the $f(R)$ theories of gravity without matter field during inflation is
\begin{eqnarray} \mathcal{A} =  \dfrac{1}{2\kappa^2}\int{d^4 x {\sqrt{-g}}f(R)}, \label{II1}\end{eqnarray}
where $f(R)$ is the function of Ricci scalar $R$,    and  we obtain  the field equation after varying  the action \eqref{II1} w.r.t. the metric tensor
\begin{eqnarray}F(R)R_{\mu\nu} - \dfrac{1}{2}f(R)g_{\mu\nu}- \nabla_\mu\nabla_\nu F(R)\nonumber \\+ g_{\mu\nu}\square F(R)= 0, \label{II2}\end{eqnarray}
where $F(R)\ = {\partial f}/{\partial R}$, $\nabla_\mu$ is the  covariant derivative and $\square \equiv \nabla_{\mu} \nabla^{\mu}$ is the  covariant D'Alembert operator.
Trace of the equation \eqref{II2} is
\begin{eqnarray} 3\square F(R) + F(R)R - 2f(R)= 0, \label{II3}\end{eqnarray}
 and equations of motion in $ f(R)$ gravity are given as
\begin{eqnarray} H^2 = \dfrac{1}{3F}\Big[\dfrac{ F R - f}{2} -3H\dot{F}\Big]; \label{II4} \end{eqnarray}
\begin{eqnarray} -2 F \dot{H} = \ddot{F} -H\dot{F}. \label{II5} \end{eqnarray}
Energy density and pressure of the universe in the $f(R)$ theory of gravity are
\begin{eqnarray} \rho = \frac{1}{\kappa^2F} \Big[\dfrac{ F R - f}{2} -3H\dot{F}\Big];  \label{II5.1}\end{eqnarray}
\begin{eqnarray} P = \frac{1}{\kappa^2F} \Big[\ddot F - \dfrac{ F R - f}{2} +2H\dot{F}\Big],  \label{II5.2}\end{eqnarray}
respectively. The slow roll conditions  \eqref{IA24} become $-\dot H/ H^2 = 1$ and $ -\ddot H/\dot H H = 1$ at the end of inflation. Putting the values  of  $F$ and $F'$ in equation \eqref{II5.1},  we have the energy density of the universe at the end of inflation as


\begin{eqnarray} \rho_{en} = \frac{3\delta(2+\delta)}{(1+\delta)}M_p^2 H^2_{en}. \label{II5.3}\end{eqnarray}
We can replace  $H_{en}$ to the $H_k$ in the \eqref{II5.3} because $H$ remains approximately constant during inflation
 \begin{eqnarray} \rho_{en} = \frac{3\delta(2+\delta)}{(1+\delta)}M_p^2 H^2_{k}. \label{II5.4}\end{eqnarray}

Putting the value of $\dot F = \dot R F^\prime$ in the equation \eqref{II4} and

 using the values  of  $F$ and $F^\prime$ where $F^\prime = \partial^2 f/ \partial R^2 $,  where prime denotes  the derivative w.r.t. $R$, we have
\begin{eqnarray}H^2 = \dfrac{\delta }{6(1+\delta)}\left[ R - 6H (1+\delta)\dfrac{\dot R}{R} \right]. \label{II8} \end{eqnarray}
We use the slow roll approximations from equations \eqref{IA24}, $ R = 6(\dot H + 2 H^2) $ and $\dot R \simeq 24H \dot H $ in the equation \eqref{II8} and obtain
\begin{eqnarray}  -\dfrac{\dot H}{ H^2} = \dfrac{1-\delta}{\delta(1+2\delta)}. \label{II9} \end{eqnarray}
Putting the value of the $-{\dot H}/{ H^2}$  from \eqref{II9} in the equation \eqref{IA8},  we obtain  the value of $\varepsilon_1$  as
\begin{eqnarray}  \varepsilon_1 = \dfrac{1-\delta}{\delta(1+2\delta)}. \label{II10} \end{eqnarray}
The parameter $\delta$ has  the range  $( \delta > (-1 +\sqrt{3})/{2}$ (i.e. $ 0.366< \delta < 1 $) from the condition of acceleration \eqref{IA9}. However,   we do not consider  $\delta > 1$  to avoid the state of super inflation \cite{Ia9}.

In the equation \eqref{IB5}, second and fourth slow roll parameters become $\varepsilon_2 = 0 $ and $ \varepsilon_4 = {\ddot F}/{H\dot F}$,  respectively,  due to the fact that  $ \dot\phi$ is absent in the Jordan frame and equation \eqref{IB6} takes the form  $ E \equiv  {3\dot F^2}/{2\kappa^2 }$  in the Jordan frame \cite{Ia9}.   Therefore,  these parameters turn out as
\begin{align}
\varepsilon_1 \equiv \dfrac{-\dot H}{H^2} ,
 &&  \varepsilon_2 = 0,
 && \varepsilon_3 \equiv \dfrac{\dot F}{2HF},
 && \varepsilon_4 \equiv \dfrac {\ddot F}{H\dot F}. \label{II12}\end{align}
We can write the equation \eqref{II5} in terms of slow roll parameters as
\begin{eqnarray} -\dfrac{\dot H}{H^2} = -\dfrac{\dot F}{2HF}\left( 1- \dfrac{\ddot F}{\dot F H}\right), \label{II13}\end{eqnarray}
\begin{eqnarray} \varepsilon_1 = -\varepsilon_3(1 - \varepsilon_4). \label{II14}\end{eqnarray}
During the slow roll inflation $ \varepsilon_4\ll 1 $  and so,  the equation \eqref{II14} changes to
\begin{eqnarray} \varepsilon_1\simeq -\varepsilon_3. \label{II15}\end{eqnarray}
We have found the scalar spectral index $n_s$ and tensor-to-scalar ratio $r$ in $ f(R)$ gravity theories in the Jordan frame after using the equation \eqref{II12},  respectively, as   \cite{Ia9, II1},
\begin{eqnarray} n_{s} \simeq 1-4\varepsilon_1 +  2 \varepsilon_3 - 2 \varepsilon_4 \label{II16}\end{eqnarray}
and
\begin{eqnarray} r = 48 \varepsilon_3^2, \label{II18}\end{eqnarray}
with  $Q_s = {E}/[{FH^2(1 + \varepsilon_3)^2}]$. We have used the Hubble parameter at the time of horizon exit at $k = a_k H_k$  as,
\begin{eqnarray} H_k = \Big( \dfrac{\pi M_p \sqrt{\mathcal{P}_s r F}}{\sqrt{2}} \Big). \label{II19}\end{eqnarray}
Clearly,   it depends on the amplitude of scalar power spectrum $\mathcal{P}_s$, tensor-to-scalar ratio $r$   and new degree of freedom $F$.
\subsection{\label{2.1}  Scalar spectral index, tensor-to-scalar ratio and reheating temperature in the Jordan frame in the $f(R)={R^{1+\delta}}/{R^\delta_c}$  model.}
In this section, we  calculate  the slow roll parameters  \eqref{II12} and power spectrum index from \eqref{I1} model. These slow roll parameters are  given by
\begin{align}
\varepsilon_1 = \dfrac{( 1- \delta )  }{ \delta(1+2\delta )};
&& \varepsilon_2 = 0;
&& \varepsilon_3 = \dfrac{(\delta - 1 )  }{(1+2\delta )};
&& \varepsilon_4 = \dfrac{(\delta - 1 )  }{\delta}.
\label{IIA1}\end{align}
We can write  $\varepsilon_3$ and $\varepsilon_4$ parameters in terms of $\varepsilon_1$ as
\begin{align}
 \varepsilon_3 = -\delta \varepsilon_1;
&& \varepsilon_4 =  -(2\delta + 1 ) \varepsilon_1.
\label{IIA2}\end{align}
We obtain  scalar spectral index by putting the value of $ \varepsilon_1$ $\varepsilon_3$ and $\varepsilon_4$ in the equation \eqref{II16} from  \eqref{IIA1}
\begin{eqnarray}n_{s} = 1- \dfrac{2( 1- \delta )^2}{ \delta(2\delta +1)}.  \label{IIA5}\end{eqnarray}
\begin{figure}[h]
\centering  \begin{center} \end{center}
\includegraphics[width=0.50\textwidth,origin=c,angle=0]{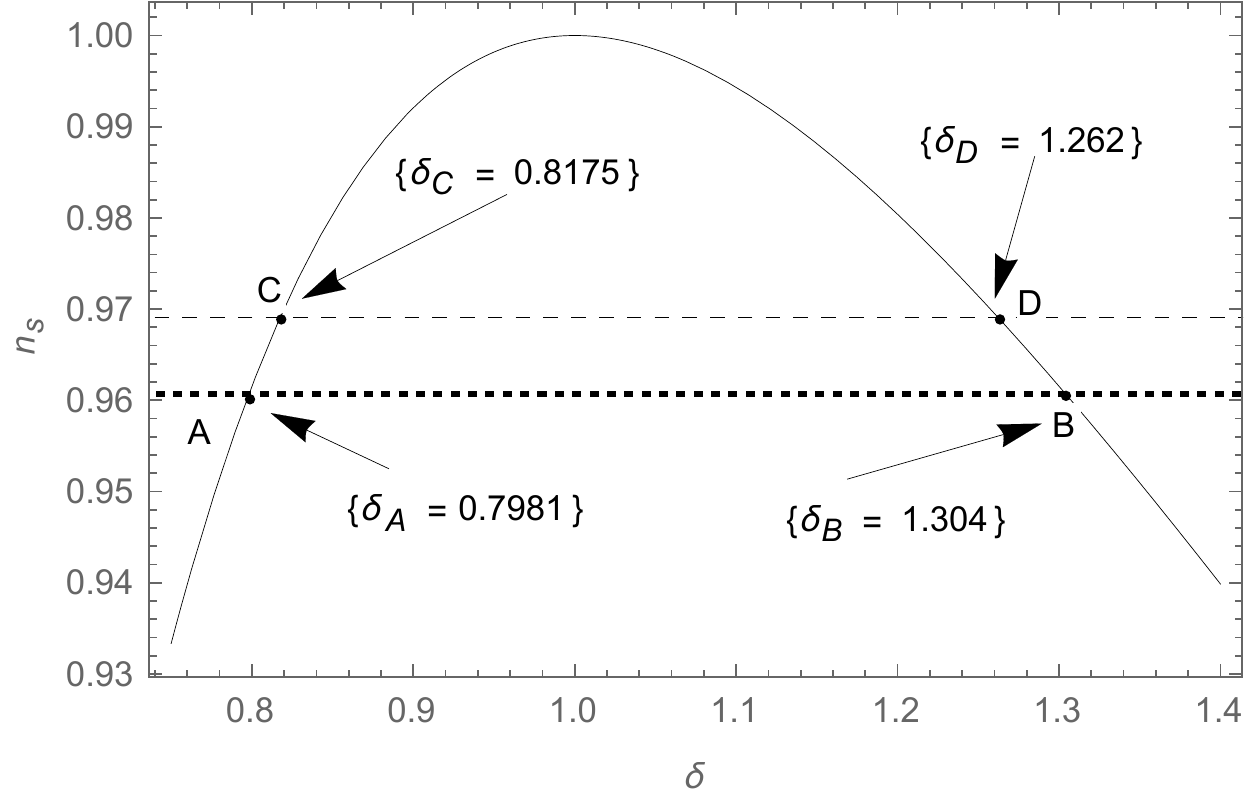}
\caption{\label{fig:p2} Plot between scalar spectral index $n_s$ and $\delta$. The  straight dashed line shows  an  observational upper limit of the $n_s = 0.9691$ and the straight dotted line is a  lower limit of $n_s = 0.9607$. In this paper,  we do not consider  the value of delta greater than one,  therefore the value of $\delta$  must  lie  between $ \delta_A\leq\delta\leq\delta_C$. As  $\delta$ increases  from $0.8175$, the value of $n_s$ crosses its upper limit.   } \label{f2}
\end{figure}
After putting the value of $ \varepsilon_3$ from the equation \eqref{IIA1} in   \eqref{II18}, we obtain  the tensor-to-scalar ratio in the Jordan frame
\begin{eqnarray} r = \dfrac{48 (1-\delta)^2}{(1+2\delta)^2}.  \label{IIA7} \end{eqnarray}
\begin{figure}[h]
\centering  \begin{center} \end{center}
\includegraphics[width=0.50\textwidth,origin=c,angle=0]{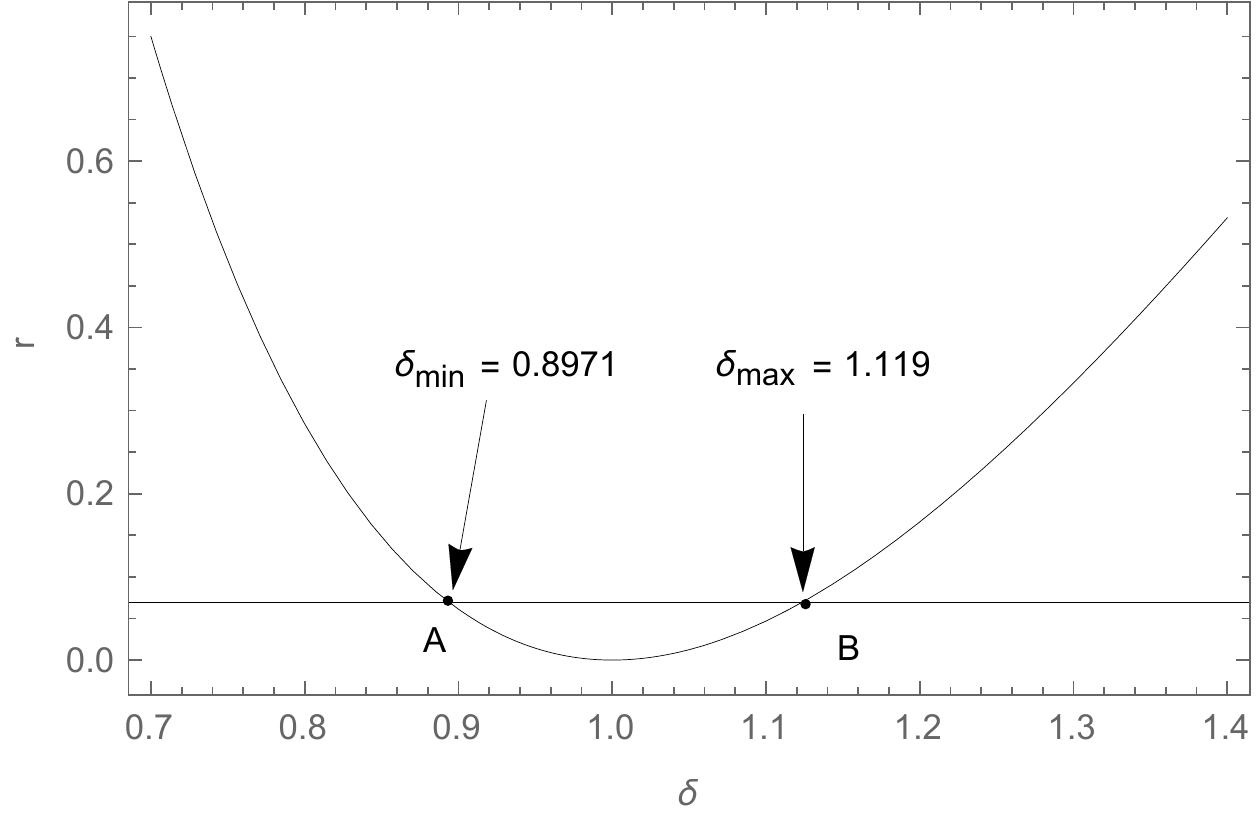}
\caption{\label{fig:p2} Plot between tensor-to-scalar ratio $r$  and $\delta$ in the Jordan frame. The straight solid  line is an  upper limit of $r = 0.065$.  This is an  observational constraint on the parameter $\delta$. Points  A and  B indicate  the minimum  and maximum value of $\delta$, implying  that  $\delta$  has  a value  $0.8971\leq\delta < 1.119$,  including   $r_{min} = 0$ at $\delta = 1$. }\label{f2}
\end{figure}
We have also found the relation between $n_{s}$ and $r$ by using equations  \eqref{IIA5} and \eqref{IIA7}
\begin{eqnarray} r \simeq -\dfrac{24\delta }{(1+2\delta)}(n_{s} -1).  \label{IIA8}\end{eqnarray}
This relation shows the dependence  on the $\delta$ parameter.
\begin{figure}[h]
\centering  \begin{center} \end{center}
\includegraphics[width=0.50\textwidth,origin=c,angle=0]{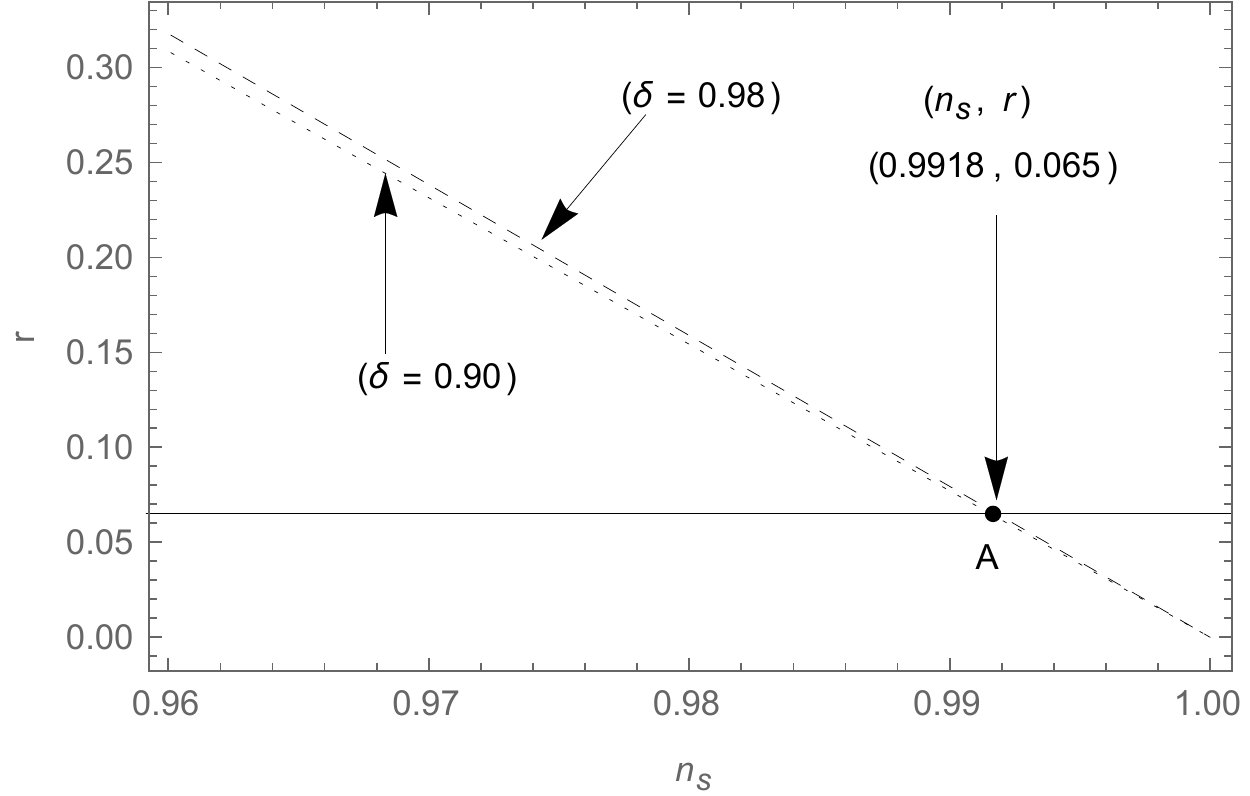}
\caption{\label{fig:p2} Plot between  tensor-to-scalar ratio $r$   and  scalar spectral index  $n_s$   for $\delta = 0.98$ and $\delta = 0.90$ in the Jordan frame.  The upper dashed line represents  $\delta = 0.98$ and  the  lower dotted  line is drawn at $\delta = 0.90$.   The  straight solid   line is an  upper bound on the value of $r$, which is $ r < 0.065 $ coming from the observations. Point A  denotes    $n_s = 0.9918$ at $r = 0.065 $.} \label{f2}
\end{figure}

\begin{figure}[h]
\centering  \begin{center} \end{center}
\includegraphics[width=0.50\textwidth,origin=c,angle=0]{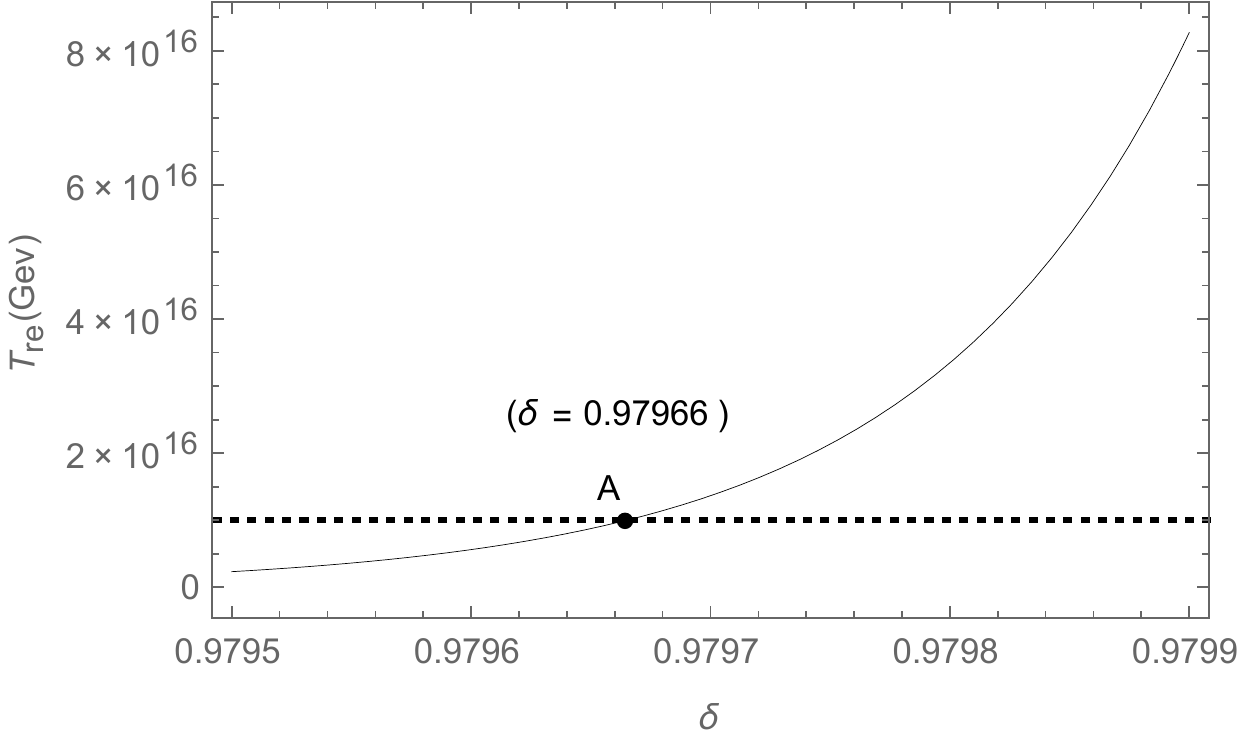}
\caption{\label{fig:p2} Plot between reheating temperature  $T_{re}$ in $GeV$ and model parameter $\delta$ in the Jordan frame.    The dotted straight line represents  the lower limit of the $T_{re} = 10^{16}$ GeV  and   intersects   the curve  at the point A  $\delta = 0.97966$.} \label{f2}
\end{figure}

We have used the relations \eqref{IIA5} and \eqref{IIA7} to obtain the  plots  of  $\delta$ vs $n_s$ and $r$.  It  can be  seen  that both  $n_s$  and $r$  depend  only  on  the model parameter  $\delta$.  Fig.1 is a plot between $\delta$ and $n_s$, showing that the allowed  values  of the parameter $\delta$ are  fixed by the observational upper and lower limit of the $n_s$. The range of the $\delta$ is $ 0.7981\leq\delta\leq 0.8175 $. The value of the $n_s$   increases  from 0.9607 to 0.9691 between the points A and C and decreases  between the points D and B.

 From  Fig.1, we can see that initially  $n_s$ increases with increasing $\delta$ and reaches  the  maximum  at $\delta = 1$ implying that  there is no tilt in the CMB power spectrum.

Fig.2 is a plot between tensor-to-scalar ratio $r$  and $\delta$.  We have a range of the $\delta$  parameter $0.8971 \leq\delta\leq 1.1192$ obtained  from the observational value of $r$.  If the value of delta  increases, the value of $r$  falls to  zero before rising again. At  $r=0$  with  $\delta = 1$, the  primordial gravitational waves  cannot be  produced. As  $r$ increases further  from  0 to 0.065  with  $\delta$  from 1 to 1.1192, having a  gravitational wave component up to its uppermost observed value  $r$  0.065, even though this $\delta$ would imply a super inflation.  Therefore,  we can set the limit of the $\delta$ as $ 0.8971\leq\delta<1$ for inflation. However, this is inconsistent with the range of $\delta$ allowed in Fig.1. It can be seen that $r<0.1$ allows a better  consistency of $r$ and  $n_s$   with respect to their variation with $\delta$.

Therefore, it is interesting  to check  it further  and  so  we  plot   tensor-to-scalar ratio  $r$  and $n_s$  in Fig.3   with two different values of the  model parameter  $\delta$. Assuming   $\delta \simeq 0.98$  gives a constraint on the value of $n_s$ and $r$. Both curves  intersect at the point (0.9918, 0.065). However, while it is satisfactory for $r$,  the value of $n_s$ is inconsistent with the observed  range discussed in Fig.1.  Again, consistency can be made stronger by allowing $r<0.1$ as shown in our model.

Reheating temperature may be determined  in terms of $\delta$ after putting   $N_k= 60$, $\rho_{en}$, and $H_k$ from equations  \eqref{II5.4} and \eqref{II19},  respectively,
\begin{eqnarray} T_{re} = \Big(\frac{11g_{s,re}}{43}\Big) \Big(\dfrac{30}{\pi^2 g_{re}} \Big) \Big(\frac{k}{a_o T_o}\Big)^3 \Big(\dfrac{\sqrt{2}}{\pi\sqrt{\mathcal{P}_s}r} \Big)\dfrac{1}{\sqrt{F}}\nonumber\\ \dfrac{3\delta(2+\delta)}{1+\delta} e^{3\sqrt{\frac{3}{2}}\frac{\delta}{1-\delta}}M_p. \label{IIA10}\end{eqnarray}
Equation \eqref{IIA10} depends on the model  parameter and the  observational parameters. By using  equation \eqref{IIA10} we can track  the behaviour of reheating temperature $T_{re}$ with $\delta$.  Thus,   $T_{re}$   also  sets a lower bound on $\delta$.   From  Fig.4,  it can be seen that  $T_{re} = 10^{16}$ GeV  sets this bound at  $\delta = 0.97966$. A lower reheating temperature would satisfy the observations  of $r$ and $n_s$ and give a more consistent value of $\delta$.

\section{\label{3} Inflationary dynamics and reheating temperature in the  Einstein frame }

Replacing   $f(R)$ by $[f(\chi) -  f(\chi),_{\chi}(R-\chi)]$ in the equation \eqref{II1} gives  action as
\begin{eqnarray} \mathcal{A} =  \dfrac{1}{2\kappa^2}\int{d^4 x {\sqrt{-g}}[f(\chi) -  f(\chi),_{\chi}(R-\chi))]}, \label{III1} \end{eqnarray}
where $\chi$ is an auxiliary field.

After  defining  $\varphi \equiv f(\chi),_{\chi}$ equation \eqref{III1}  can be  written as
\begin{eqnarray}\mathcal{A} =  \int{d^4 x {\sqrt{-g}}\left[\dfrac{1}{2\kappa^2}\varphi R -  U(\varphi)\right]} ,\label{III2}\end{eqnarray}
where $U(\varphi)$ is  potential of the field $(\varphi)$  as
\begin{eqnarray} U(\varphi)\equiv \dfrac{\chi(\varphi)\varphi - f(\chi(\varphi))}{2\kappa^2}. \label{III3}\end{eqnarray}
We can rewrite the equation \eqref{II1} as
\begin{eqnarray}\mathcal{A} = \int d^{4}x\sqrt{-g}\left(\frac{1}{2\kappa^{2}}f(R)R - U \right),  \label{III4} \end{eqnarray}
where
\begin{eqnarray}U = \dfrac{FR - f}{2\kappa^2}.   \label{III5}\end{eqnarray}
Invoking  a  conformal transformation of the metric tensor
$\tilde{g}^{\mu\nu} = \Omega^2 g^{\mu\nu}$ and $\sqrt{-g}= \Omega^{-4}\sqrt{-\tilde{g}} $, we obtain  the action in the Einstein frame ($\mathcal{A}_E$) as

\begin{widetext}
\begin{eqnarray}\mathcal{A}_E = \int d^{4}x\sqrt{-\tilde{g}}\left[\frac{1}{2\kappa^{2}}F\Omega^{-2}( \tilde{R} + 6\tilde{\square}\omega -6 \tilde{g}^{\mu\nu}\partial_{\mu}\omega\partial_{\nu}\omega) - \Omega^{-4}U \right], \label{III6}\end{eqnarray}
\end{widetext}
where $\omega\equiv\ln\Omega $,    $ \partial_{\mu}\omega\equiv{\partial\omega}/{\partial\tilde{x}^{\mu}}$,   $\tilde{\square}\omega\equiv ({1}/{\sqrt{-g}})\partial_{\mu}(\sqrt{-\tilde{g}}\tilde{g}^{\mu\nu}\partial_{\nu}\omega)$,\\
$R = \Omega^2( \tilde{R} + 6\tilde{\square}\omega -6 \tilde{g}^{\mu\nu}\partial_{\mu}\omega\partial_{\nu}\omega)$.

We can rewrite the equation \eqref{III6}  by  redefining  the scalar field
$ \kappa\phi = \sqrt{3/2}\ln F $ and $ \Omega^2  = F $  in the Einstein frame as
\begin{eqnarray}
\mathcal{A}_E = \int d^{4}x\sqrt{-\tilde{g}}\left[\frac{1}{2\kappa^{2}}\tilde{R}-\dfrac{1}{2}\tilde{g}^{\mu\nu}\partial_{\mu}\phi\partial_{\nu}\phi - V(\phi)\right]
\label{III7}\end{eqnarray}
where
\begin{eqnarray}V(\phi)= \dfrac{U}{F^2} = \dfrac{FR - f}{2\kappa^2F^2}. \label{III8}\end{eqnarray}
Equation \eqref{III7} shows that the scalar field $\phi$ is minimally coupled with the curvature. Since action \eqref{III7} is same for the canonical single scalar field,   therefore the dynamical equations and the conclusions drawn  are equivalent in the Einstein frame for the slow roll inflation.

We have evaluated the potential of the scalar field $V(\phi)$ from the equation \eqref{III8} by using the value of $R$, $f(R)$ and $F$ in term of the scalar field, $\phi$  as \cite{I20}
\begin{eqnarray} V(\phi) = \dfrac{ \delta R_c M_p^2}{2 (1+\delta)^{\frac{1+\delta}{\delta}} }\left[ e^{\sqrt{\frac{2}{3} }\frac{\phi}{M_p}}\right]^{\frac{1-\delta}{\delta}}.  \label{III9}\end{eqnarray}
\begin{figure}[h]
\centering  \begin{center} \end{center}
\includegraphics[width=0.50\textwidth,origin=c,angle=0]{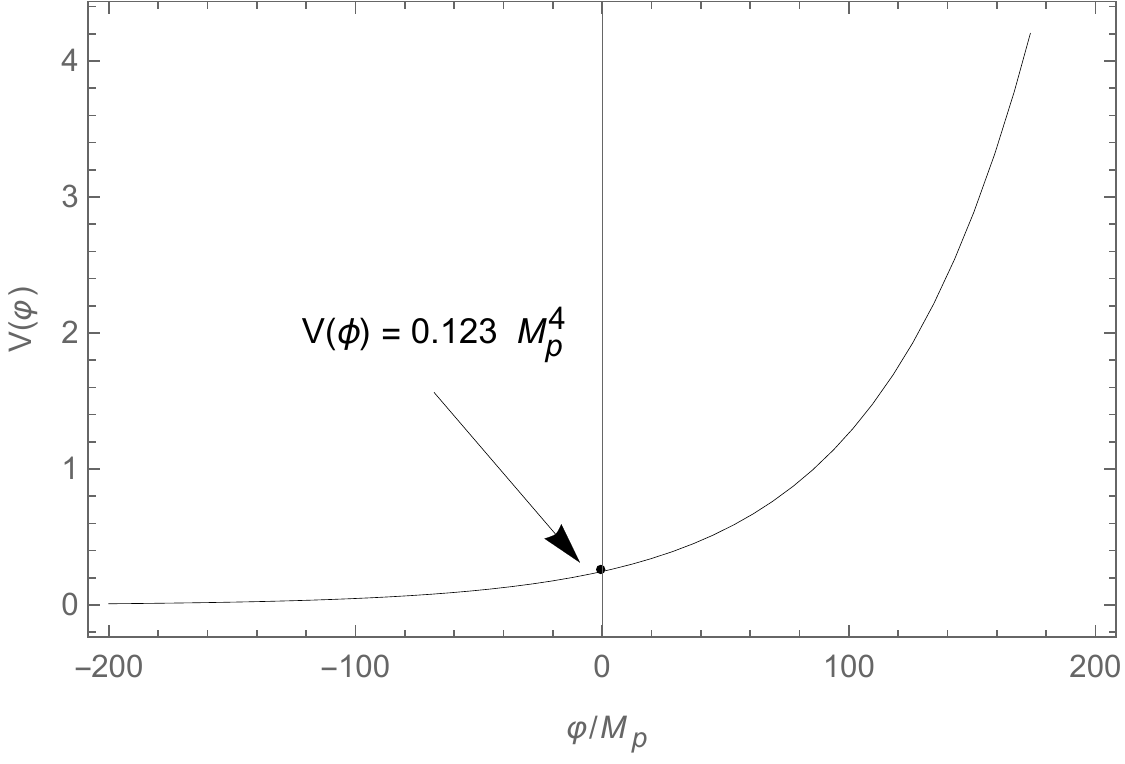}
\caption{\label{fig:p2} Plot showing  the behavior of the scalar field potential $V(\phi)$ with respect to  scalar field $\phi/M_p$. Here, we consider  $\delta = 0.98$ and $R_c = 1$ in $M_p^2$ units. Homogeneous and isotropic scalar field potential has minimum if $\phi\rightarrow \infty$. Inflation begins  at $\phi= M_p$ and rolling down towards  $\phi = 0$.} \label{f2}
\end{figure}
Interestingly, potentials of this type  are able to produce slow roll inflation in the very early universe.

In Fig. 5, slow roll inflation ends as  $\phi\rightarrow \infty$. Such a potential has  a minima at $\infty$,  therefore the  scalar field (inflaton field) does  not oscillate about minima. It goes on  rolling towards  infinity and decays into other particles in  reheating phase. In this   mechanism,   potential has  a global minima at the $\phi_{min}$  and  the inflaton field  oscillates  about it.  It may  decay by a direct coupling to the matter field and other scalar field. Alternatively,  reheating process may be followed by gravitational particle production \cite{VI1,Ia3}. However,  in  the present   paper,  we do not intend to discuss  the process of the particle production during reheating.

Slow roll parameters in the scalar-tensor theory are defined by the equation \eqref{IB5} but in the Einstein frame  $ F = 1$,  and so  $\tilde{\varepsilon}_3$  and $\tilde{\varepsilon}_4$  vanish \cite{Ia9}.  Equation \eqref{IB5} becomes
\begin{align} \tilde{\varepsilon}_1 \equiv -\dfrac{\dot {\tilde{H}}}{\tilde{H}^2};
&& \tilde{\varepsilon}_2\equiv\dfrac{\ddot \phi}{\tilde{H}\dot\phi};
&&\tilde{\varepsilon}_3 = 0;
&&\tilde{\varepsilon}_4 = 0  \label{III10} \end{align}
where $\dot\phi= {\partial\phi}/{\partial \tilde{t}}$ and $\ddot\phi = {\partial^2\phi}/{\partial \tilde{t}^2}$,  $ d\tilde{t} = \sqrt{F}dt$, $ \tilde{a} = \sqrt{F}a $ and the Hubble parameter  $\tilde{H} \equiv(1/\tilde{a}){d\tilde{a}}/{d\tilde{t}} = ({1}/\sqrt{F})\left( H + {\dot{F}}/{2F}  \right)$.

There are two parameters $\tilde{\varepsilon}_1$ and $\tilde{\varepsilon}_2$ surviving  in the Einstein frame. These  can be expressed in terms of potential as well as in terms of $f(R)$  as   \cite{III1}
\begin{eqnarray} \tilde{\varepsilon}_1 \equiv \dfrac{1}{2\kappa^2}\left(\dfrac{V,_\phi}{V}\right)^2, \label{III11}\end{eqnarray}
\begin{eqnarray}\tilde{\varepsilon}_1 = \dfrac{1}{3}\left(\dfrac{2f - FR}{FR - f} \right)^2, \label{III12}\end{eqnarray}
\begin{eqnarray}\tilde{\varepsilon}_2 = \dfrac{1}{2\kappa^2}\left(\dfrac{V,_\phi}{V}\right)^2 - \dfrac{1}{\kappa^2}\left(\dfrac{V,_{\phi\phi}}{V}\right),  \label{III13}\end{eqnarray}

\begin{eqnarray} \tilde{\varepsilon}_2 =\tilde{\varepsilon}_1 - \eta, \label{III14}\end{eqnarray}
where $\eta$ is given by the equation \eqref{IA34}.

Further,   $\eta$ can be expressed  in terms of the $f(R)$   as   \cite{III1}
\begin{eqnarray}\eta = \dfrac{2}{3}\dfrac{F^2}{F^\prime(FR-f)}- \dfrac{2RF}{(FR-f)} + \dfrac{8}{3} \label{III15}\end{eqnarray}

 Using the slow roll condition \eqref{IA33.1}  in the $\tilde{\rho}_\phi \simeq (1/2){\dot{\tilde{\phi}}}^2 + V(\phi)$, the energy density of the scalar field becomes $\tilde{\rho}(\phi)\simeq V(\phi)$ while  the Friedmann's equation  at the time of $k = \tilde{a}_k \tilde{H}_k$   turns out to be   $ \tilde{H}^2_k \simeq \tilde{\rho}_k(\phi)/3 M_p^2$. At the end of the inflation,  $(1/2)\dot\phi^2 \simeq V(\phi)$ and the energy density $\tilde{\rho}_{en}(\phi)$ becomes
\begin{eqnarray} \tilde{\rho}_{en}(\phi)\simeq \dfrac{3}{2}V_{en}(\phi). \label{III15.1}\end{eqnarray}
Inflaton potential at the end of inflation is
\begin{eqnarray} V_{en}(\phi) = V_k(\phi) e^{\sqrt{\frac{2}{3}}\Big[\frac{\phi_k}{M_p} - \frac{\phi_{en}}{M_p}  \Big]\frac{1-\delta}{\delta}}. \label{III15.2}\end{eqnarray}
Here,  we have  assumed that $ \phi_{en} = 0$;  $\phi_k = M_p $ and $V(\phi)_k = 3\tilde{H}^2_k M_p^2$. Putting these values in the above equation, we obtain   \begin{eqnarray}  V_{en}(\phi) = 3\tilde{H}^2_k M_p^2 e^{\sqrt{\frac{2}{3}}\Big( \frac{\delta-1}{\delta}\Big)}. \label{III15.3}\end{eqnarray}

Using the above value of the $V(\phi)_{en}$ in the \eqref{III15.1}, $\tilde{\rho}_{en}(\phi)$  becomes
\begin{eqnarray}  \tilde{\rho}_{en}(\phi)\simeq \frac{9}{2}M_p^2 \tilde{H}^2_k  e^{\sqrt{\frac{2}{3}}\Big( \frac{\delta-1}{\delta}\Big)}. \label{III15.4}\end{eqnarray}
In the Einstein frame, Hubble  parameter  in terms of $\tilde{\mathcal{P}}_s$ and $\tilde{r}$  is given as,
\begin{eqnarray} \tilde{H}_k  = \Big(\dfrac{\pi M_P \sqrt{\tilde{\mathcal{P}}_s}\tilde{r}}{\sqrt{2}} \Big), \label{III15.5} \end{eqnarray}
at the horizon exit.

Scalar spectral index $\tilde{n}_s$ in the Scalar-Tensor theory is  given by the equation \eqref{IB8} \cite{Ib1}.  Thus,  using the values  of $ \tilde{\varepsilon}_1$, $ \tilde{\varepsilon}_2$,  $\tilde{\varepsilon}_3$ and $\tilde{\varepsilon}_4$ from the equation \eqref{III10},

\begin{eqnarray} \tilde{n}_{s}-1\simeq -4\tilde{\varepsilon}_1 -2\tilde{\varepsilon}_2. \label{III16}\end{eqnarray}
which is further given as (from the equation \eqref{III14})
\begin{eqnarray} \tilde{n}_{s}\simeq 1 -6\tilde{\varepsilon}_1 + 2\eta . \label{III17}\end{eqnarray}

Further, we have found the tensor-to-scalar ratio by putting the value of $Q_s = {\dot\phi^2}/{\tilde{H}^2}$ in the equation \eqref{IB9} in the Einstein frame \cite{Ia9} as
\begin{eqnarray} \tilde{r}\simeq -16\frac{\dot{\tilde{H}}}{\tilde{H}^2} = 16\tilde{\varepsilon}_1. \label{III18}\end{eqnarray}
We can also  determine the   number of e-foldings in the Einstein frame from the equation \eqref{IA37} as
\begin{eqnarray} N = \int^{\phi_{60}}_{\phi_{end}}{\dfrac{d(\phi/M_p)}{\sqrt{2\tilde{\varepsilon}_1}} }. \label{III19}\end{eqnarray}
This expression is further  used to calculate the value of the $\delta$ parameter.

\subsection{\label{3.1} Scalar spectral index, tensor-to-scalar ratio and reheating temperature in the Einstein frame in the $f(R) ={R^{1+\delta}}/{R^\delta_c}$  model. }

In this section we  calculate  $\tilde{\varepsilon}_1$ and $\eta$ in  the  model \eqref{I1} by using the equation \eqref{III12} and \eqref{III15},  respectively, as
\begin{eqnarray} \tilde{\varepsilon}_1 = \dfrac{1}{3}\dfrac{(1- \delta)^2}{\delta^2}  \label{IIIA1}\end{eqnarray}
and
\begin{eqnarray}\eta = \dfrac{2(1-\delta)^2}{3\delta^2} = 2\tilde{\varepsilon}_1.  \label{IIIA2}\end{eqnarray}
These expressions  show  dependence  on the single parameter $\delta$ only, and thus  are sensitive to the $f(R)$ model parameter.

This leads us to  obtain  the expressions for scalar spectral index $\tilde{n}_{s}$ and tensor-to-scalar ratio $\tilde{r}$  using the value of  $\tilde{\varepsilon}_1$ and $\eta$ in the equation \eqref{III17} and \eqref{III18},  respectively, as
\begin{eqnarray} \tilde{n}_{s} \simeq 1-\dfrac{2}{3}\left(\dfrac{1-\delta}{\delta}\right)^2,  \label{IIIA3}\end{eqnarray}
\begin{figure}[h]
\centering  \begin{center} \end{center}
\includegraphics[width=0.50\textwidth,origin=c,angle=0]{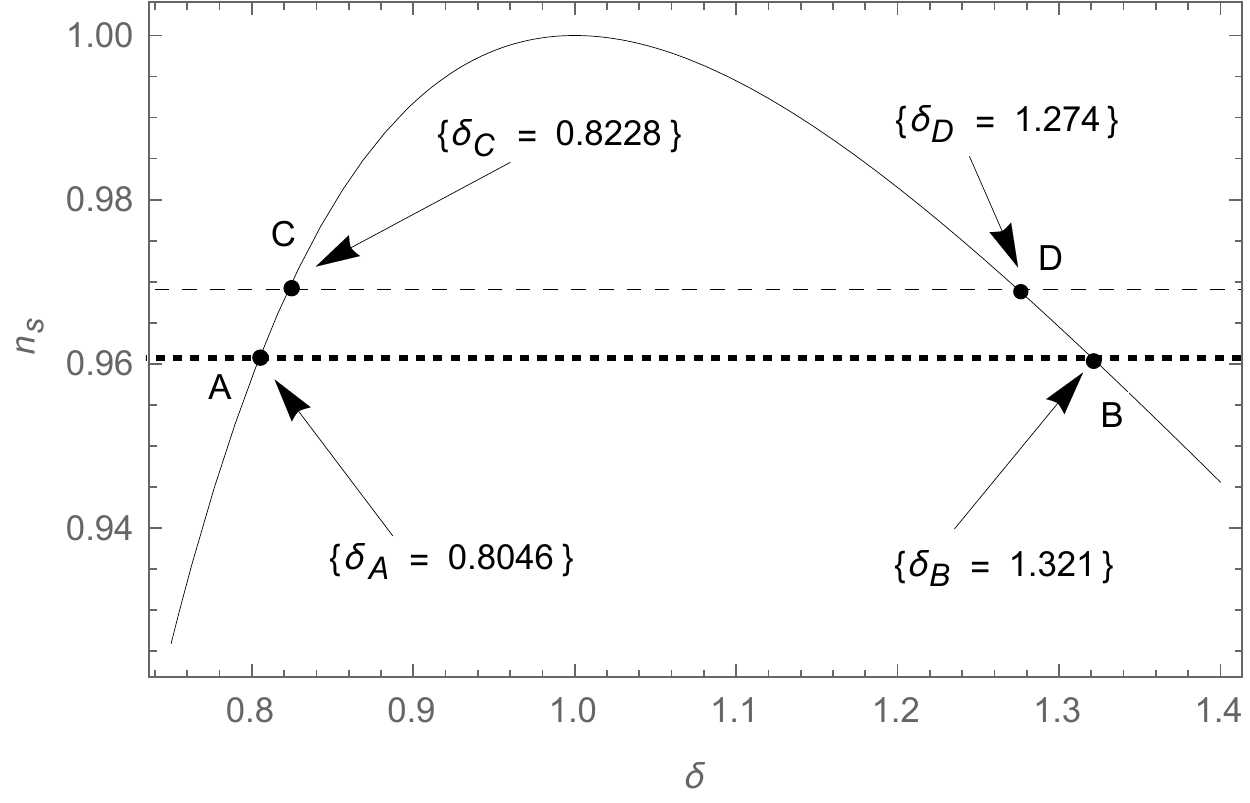}
\caption{\label{fig:p2} Plot between scalar spectral index $\tilde{n}_s$ and $\delta$ parameter in the Einstein frame. The dashed straight line is the  upper limit of the $\tilde{n}_s = 0.9691$ and dotted straight line is the   lower limit $\tilde{n}_s = 0.9607$. The value of $\delta$ lies over the range  $0.8046\leq\delta\leq0.8228$. The variation also  shows a  maximum  at $\delta = 1$ and $\tilde{n}_s = 1$.} \label{f2}
\end{figure}

\begin{eqnarray} \tilde{r} =\dfrac{16}{3} \dfrac{(1-\delta)^2}{\delta^2}. \label{IIIA4}\end{eqnarray}
Using  equations \eqref{IIIA3} and \eqref{IIIA4},    we get  the relation between spectral index  and tensor-to-scalar ratio as
\begin{eqnarray} \tilde{r} = -8 (\tilde{n}_s - 1). \label{IIIA5}\end{eqnarray}
It may be noticed that the  above relation is independent of  $\delta $ parameter.\\
\begin{figure}[h]
\centering  \begin{center} \end{center}
\includegraphics[width=0.50\textwidth,origin=c,angle=0]{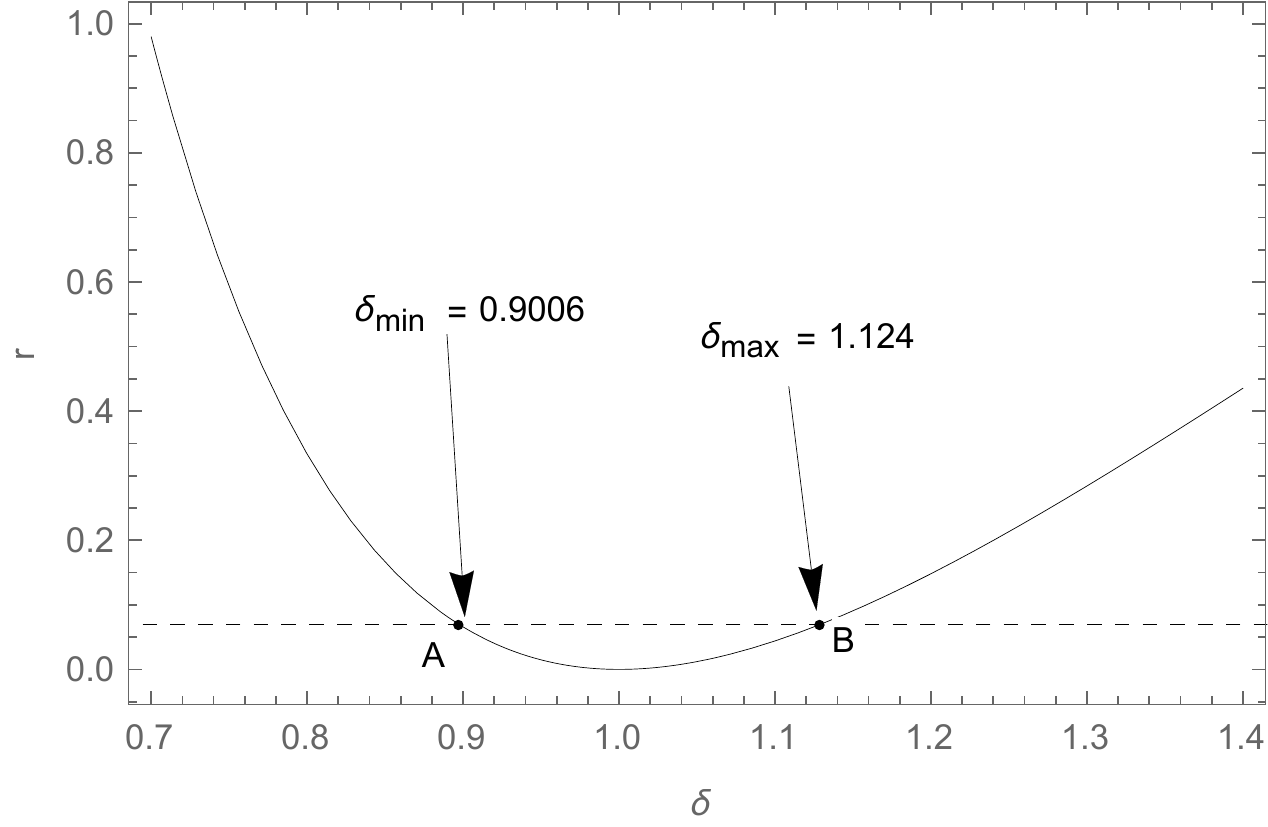}
\caption{\label{fig:p2} Plot between tensor-to-scalar ratio $\tilde{r}$ and $\delta$ in the Einstein frame. The dashed straight line at $\tilde{r} = 0.065$  sets the observational limit on the $\delta$ parameter. It marks a  minimum  at $\delta=1$ with  $r_{min} = 0$.} \label{f2}
\end{figure}
Further,   we  calculate the value of $\delta $ from the number of e-foldings  by putting the value of $\tilde{\varepsilon}_1$  in the equation \eqref{III19},
\begin{eqnarray}  N = \int^{\phi_{60}}_{\phi_{end}}{\dfrac{d(\phi/M_p)}{\sqrt{2(1-\delta)^2/3\delta^2}} }. \label{IIIA6}\end{eqnarray}
Integrating  equation \eqref{IIIA6},
\begin{eqnarray}  N = \sqrt{\dfrac{3}{2}}\dfrac{\delta}{(1-\delta)}\left[\dfrac{\phi_{60}}{M_p} - \dfrac{\phi_{end}}{M_p} \right]. \label{IIIA7}\end{eqnarray}
If  we assume that inflation starts  at $\phi_{60}= M_p$ and ends  at  $ \phi_{end}=0$, then  from  equation \eqref{IIIA7} we have
\begin{eqnarray}  N = \sqrt{\dfrac{3}{2}}\dfrac{\delta}{(1-\delta)} \label{IIIA8}\end{eqnarray}
We have kept the  number of e-folds $  N = 60$  with a view to solve the horizon problem and flatness problem in the $\Lambda$CDM model.    Putting the value of $ N$ in the  equation \eqref{IIIA8} we can constrain our model parameter.  Thus,
\begin{eqnarray}  \dfrac{\delta}{(1-\delta)} = 60\sqrt{\dfrac{2}{3}} \label{IIIA9} \end{eqnarray}
\begin{eqnarray} \delta  = 0.97999\simeq 0.98  \label{IIIA10}\end{eqnarray}
which is very close  to 1.    It is particularly significant because using the value of  $\delta$, we are able to  calculate the  scalar spectral index  and tensor-to-scalar ratio, both  in the Jordan frame and the Einstein frame.

\begin{figure}[h]
\centering  \begin{center} \end{center}
\includegraphics[width=0.50\textwidth,origin=c,angle=0]{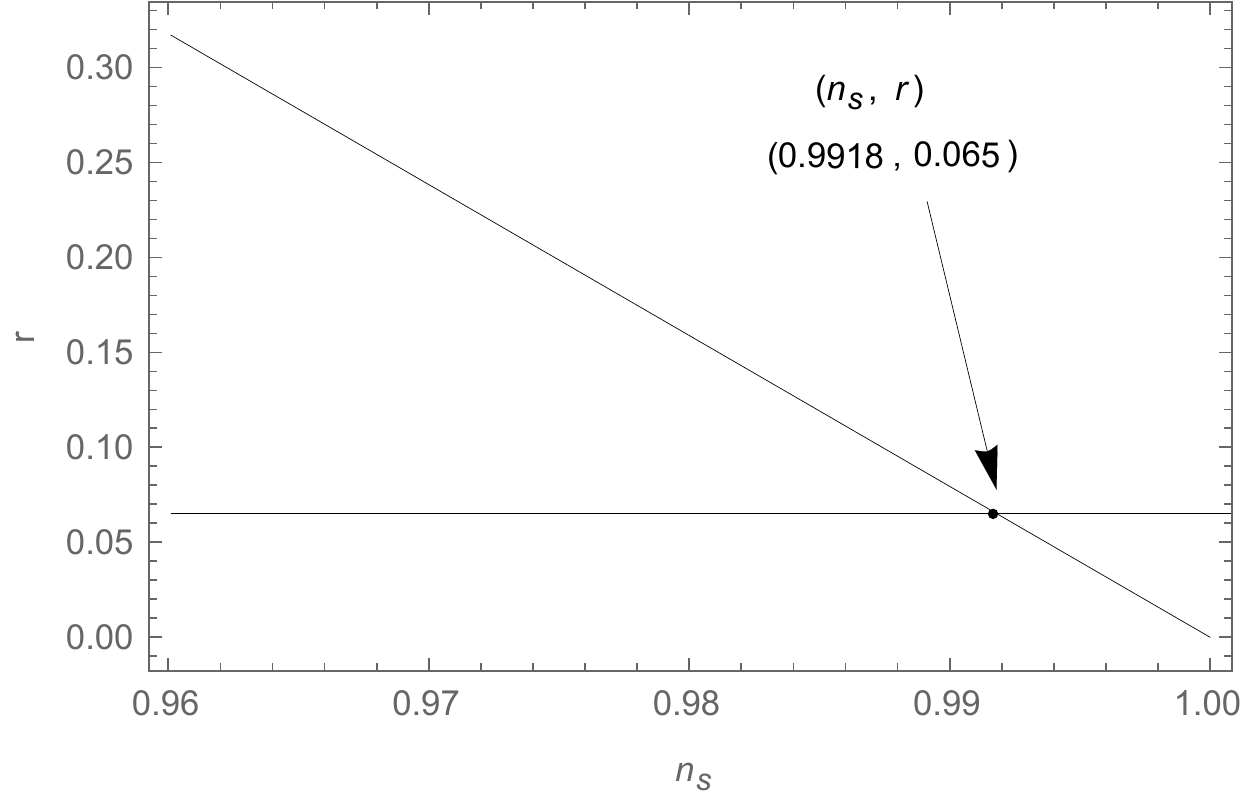}
\caption{\label{fig:p2} Plot between tensor-to-scalar ratio $\tilde{r}$ and scalar spectral index $ \tilde{n}_s $ for  $\delta= 0.98$ in the Einstein frame. The horizontal  straight line is an  upper bound on the $\tilde{r}$. At the point of intersection  A,   the value of $\tilde{n}_s$ is $ 0.9918$.}  \label{f2}
\end{figure}

\begin{figure}[h]
\centering  \begin{center} \end{center}
\includegraphics[width=0.50\textwidth,origin=c,angle=0]{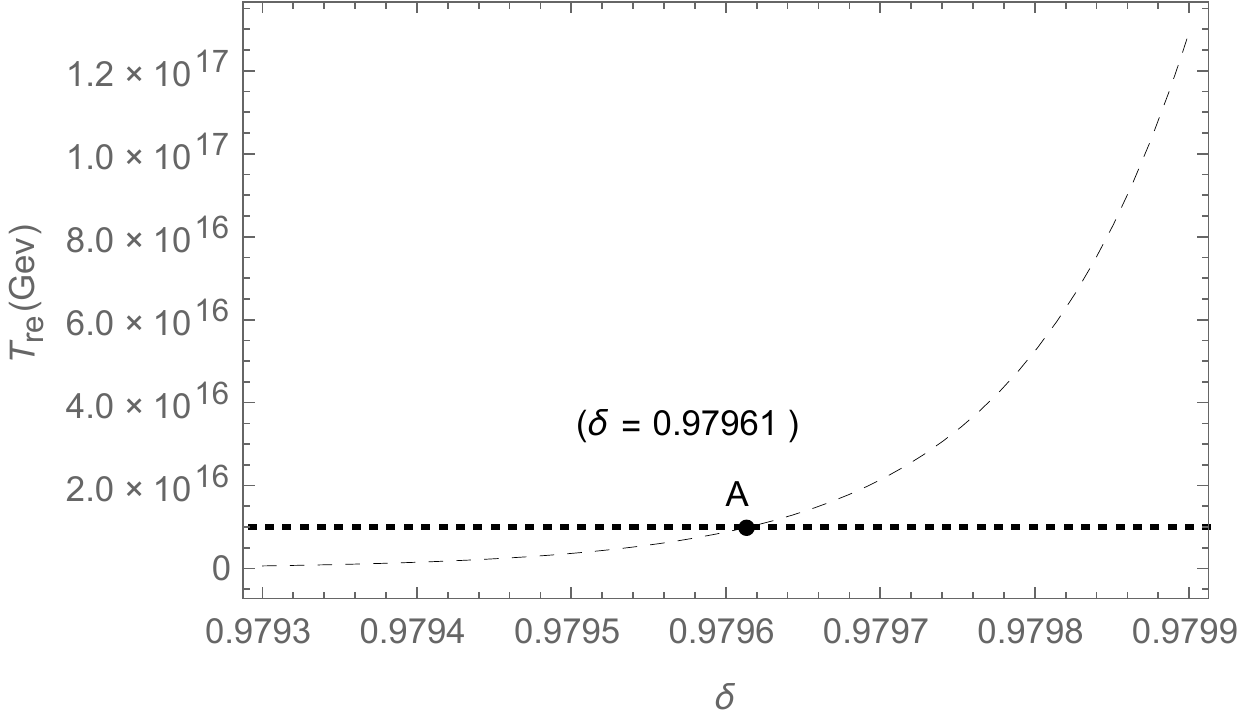}
\caption{\label{fig:p2} Plot between the reheating temperature  $T_{re}$ and the  model  parameter $\delta$ in the Einstein frame.  The straight dotted line provides  a lower bound on the reheating temperature giving  $\delta = 0.97961.$  } \label{f2}
\end{figure}
The equation \eqref{IIIA3}  gives the relation  between $n_s$ and $\delta$ also  seen  in  Fig.6,   where   we have constrained  the value of $\delta$ parameter from  the observational upper and lower limits  of the $n_s$. The  admissible range is  $ 0.8046\leq\delta\leq 0.8228 $. The  solid curve has a maximum at the value of $\tilde{n}_s = 1$ and $\delta = 1$   which implies that  there is no tilt in the power spectrum of the CMB.

In  Fig.7, we have shown  a range of the parameter $0.9006 \leq\delta\leq 1.124$ from the observational value of $\tilde{r}$. If the value of delta  increases  then the  $\tilde{r}$ goes towards the zero before  increasing  with  $\delta$. The value of $\tilde{r}$ is zero at $\delta = 1$,  which  means that  the primordial gravitational waves  can not be produced  during  inflation but metric fluctuations must  present in the early universe. Since  an increasing  value of $\delta$  makes  $\tilde{r}$ become greater than 0.065,  therefore,  we can set the optimal bound  on the $\delta$  as  $ 0.9006\leq\delta<1$  for inflation.

Indeed, if  we  satisfy  $r<0.065$ then  $\tilde{n}_s$  must be at least equal to    $0.9918$, which can be further  relaxed if $r<0.1$.   This behaviour  can be seen  from  Fig.8.

With the above analysis, we can comment   that although there is a lack of complete concordance  range of the $\delta$ parameter from the Fig.6 and Fig.7, however,  from the Fig.3 and Fig.8, we have obtained  the values  of  $n_s\leq 0.9918$ with  our model parameter  $\delta = 0.98$.  Therefore,   we may conclude that   $\delta$  lies in the range  $0.8046 \leq \delta \leq 1$ in the Einstein frame.

Now, the  reheating temperature from \eqref{IC15} given in the Einstein frame after
putting the value of $\tilde{N}_k$, $\tilde{H}_k $ and $\tilde{\rho}_{\phi}(en)$ from equations \eqref{IIIA8}, \eqref{III15.5} and \eqref{III15.4},  respectively,  into \eqref{III19} becomes as
\begin{eqnarray} \tilde{T}_{re} = \frac{9}{2} \Big(\frac{11g_{s,re}}{43}\Big) \Big(\dfrac{30}{\pi^2 g_{re}} \Big) \Big(\frac{k}{a_o T_o}\Big)^3 \Big(\dfrac{\sqrt{2}}{\pi \sqrt{\tilde{\mathcal{P}}_s}\tilde{r}} \Big)\nonumber\\ e^{\sqrt{\frac{3}{2}}\frac{\delta-1}{\delta}} e^{3\sqrt{\frac{2}{3}}\frac{\delta}{1-\delta}} M_p . \label{IIIA12}\end{eqnarray}
The  behaviour of  $\tilde{T}_{re}$ from \eqref{IIIA12} with respect to   $\delta$ can be seen in the Fig.9,  where   $\tilde{T}_{re} = 10^{16}$  GeV is a lower bound indicating  $\delta = 0.97961$. This provides  the minimum energy  required to decouple gravitational field from other fields.

\section{\label{4} Are the Jordan frame and the Einstein frame equivalent? }

In the foregoing discussion,  we have seen  conformal equivalence   with the help of $n_s$, $r$ and $T_{re}$ between the Jordan and the Einstein frame in the very early universe.  In the Jordan frame,   we obtained  scalar spectral index and tensor-to-scalar ratio, respectively,  as
\begin{align}
  n_s = 1 -  \dfrac{2}{\delta}\dfrac{( 1- \delta )^2  }{ (1+2\delta)};
&& r = 48 \dfrac{(1 - \delta)^2}{(1+2\delta)^2}
,\label{IV1}\end{align}

and  in the Einstein frame as
\begin{align}
 \tilde{n}_s =  1 -  \dfrac{2}{3}\dfrac{( 1- \delta )^2  }{ \delta^2};
&& \tilde{r} =  \dfrac{16}{3}\dfrac{(1-\delta)^2}{\delta^2}.
\label{IV2}\end{align}
\begin{figure}[h]
\centering  \begin{center} \end{center}
\includegraphics[width=0.50\textwidth,origin=c,angle=0]{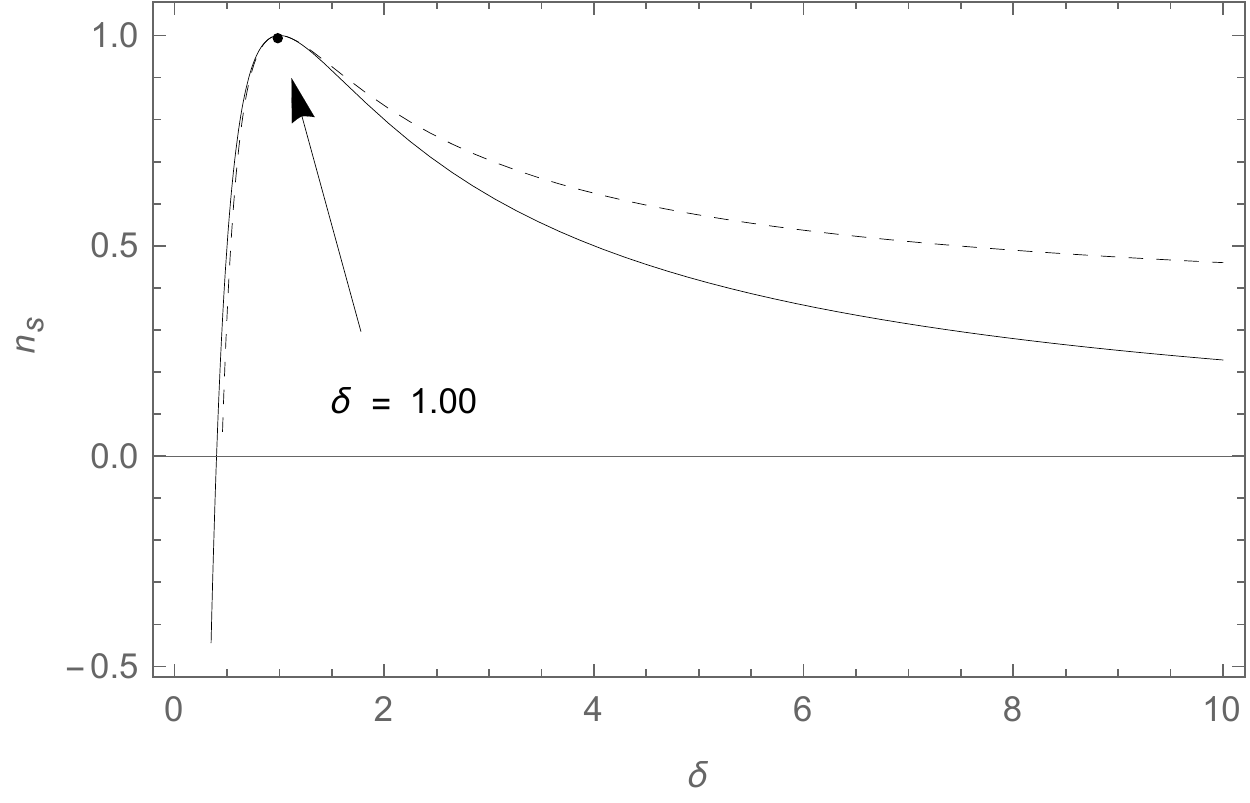}
\caption{\label{fig:p2}  Plot between  scalar spectral index $ n_s $ and $\delta$ in the  Jordan frame (solid curve) and the Einstein frame  (dashed  curve).    As  $\delta$ rises above  $\delta=1$,  we observe a marked difference between the two frames,  while below $\delta=1$ they are approximately  indistinguishable from each other. } \label{f2}
\end{figure}

\begin{figure}[h]
\centering  \begin{center} \end{center}
\includegraphics[width=0.50\textwidth,origin=c,angle=0]{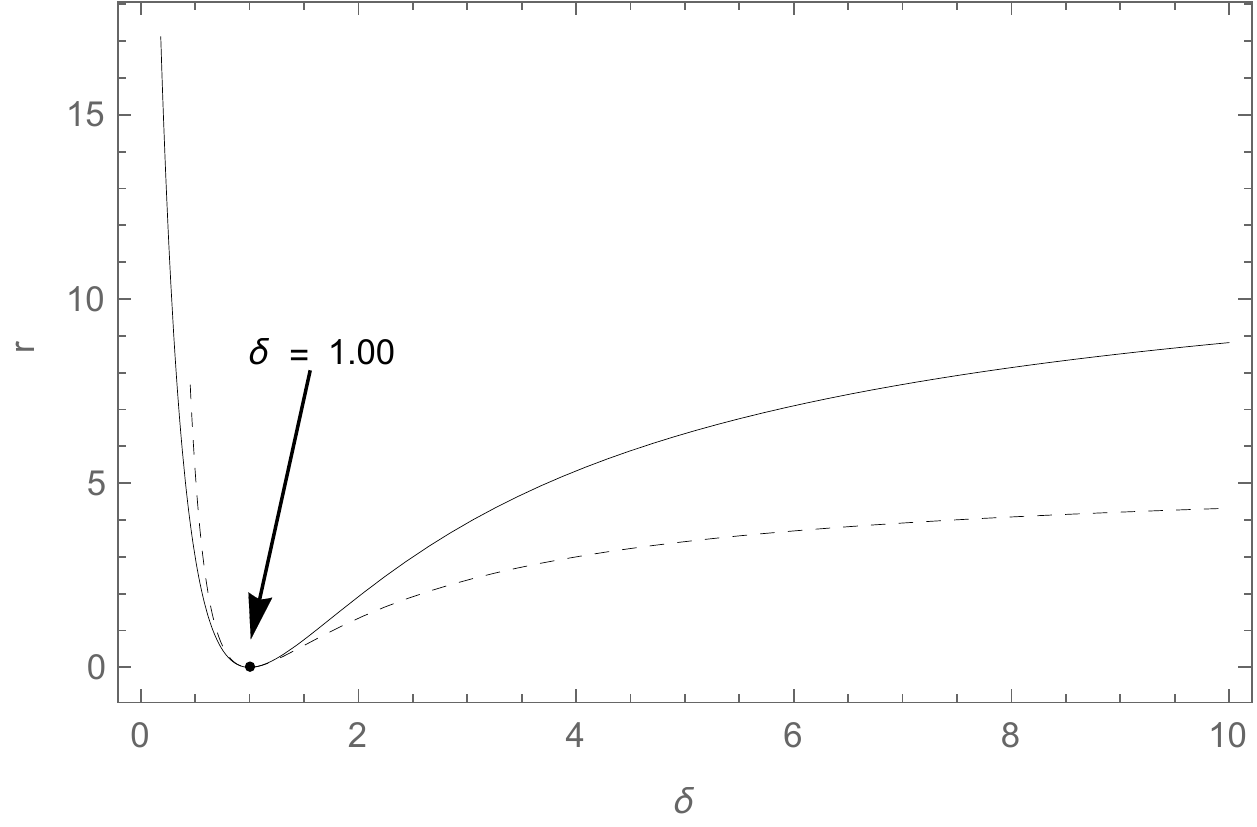}
\caption{\label{fig:p2} Plot between  tensor-to-scalar ratio $ r $ and $\delta$ in the Jordan frame (solid curve) and the Einstein frame (dashed  curve).} \label{f2}
\end{figure}

\begin{figure}[h]
\centering  \begin{center} \end{center}
\includegraphics[width=0.50\textwidth,origin=c,angle=0]{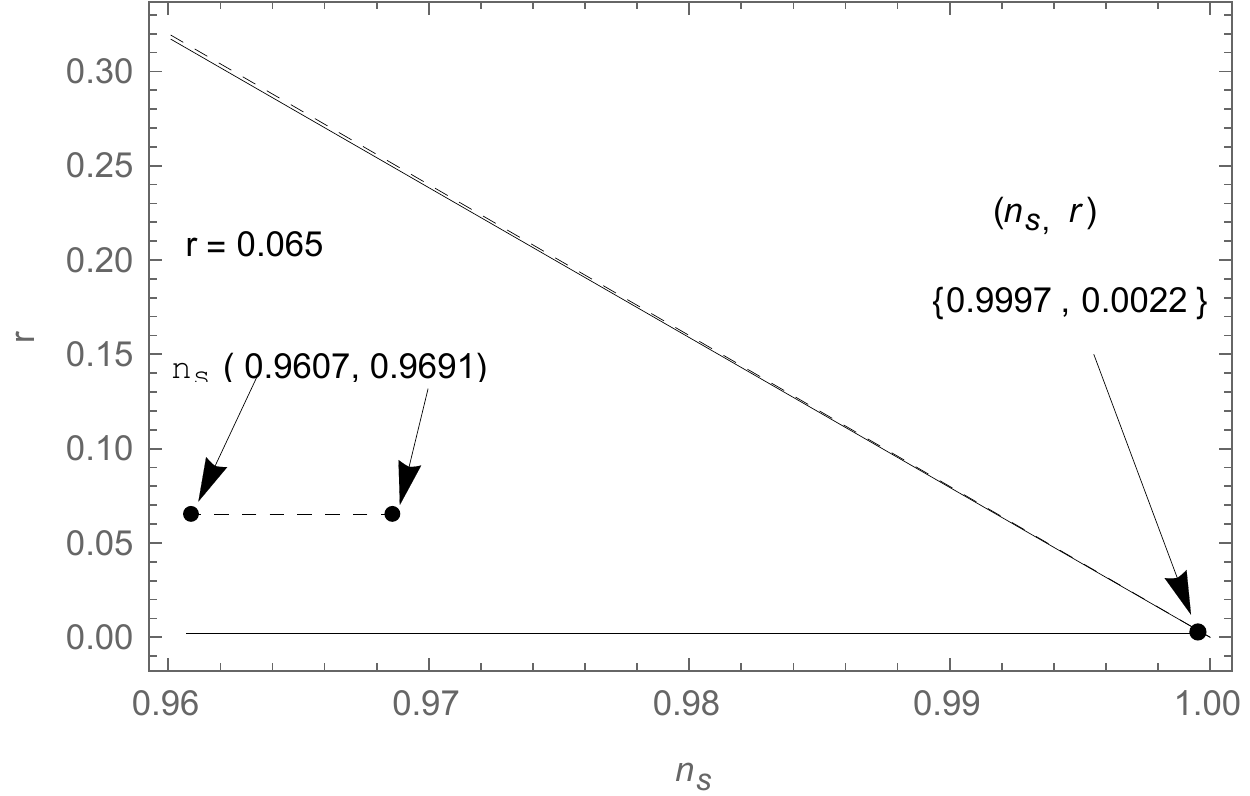}
\caption{\label{fig:p2}  Plot between  tensor-to-scalar ratio $ r $ and scalar spectral index $n_s$ at $\delta = 0.98$  in the Jordan frame (solid curve) and the Einstein frame  (dashed curve).  The horizontal  dashed line provides  an observational limit   on $n_s$ at $r = 0.065$,   whereas   the  lower solid horizontal  straight line is drawn by using the calculated value of $r = 0.0022$.} \label{f2}
\end{figure}

\begin{figure}[h]
\centering  \begin{center} \end{center}
\includegraphics[width=0.50\textwidth,origin=c,angle=0]{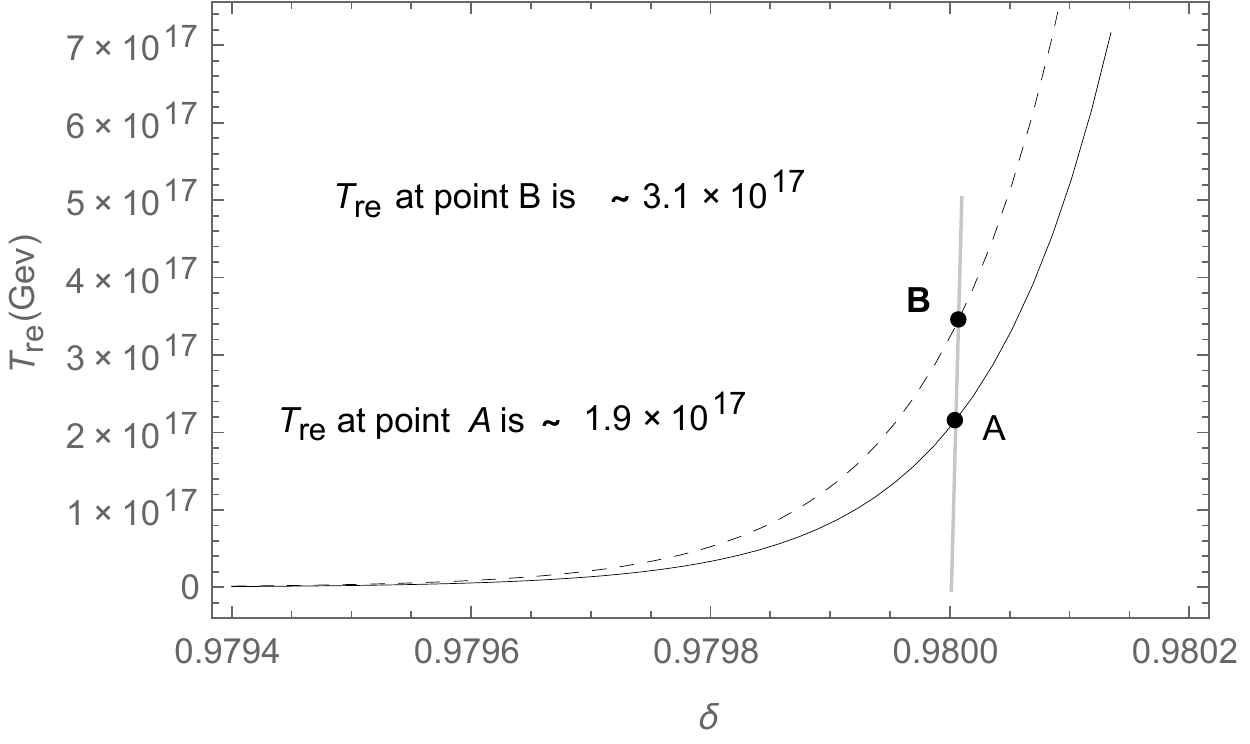}
\caption{\label{fig:p2} Plot between reheating temperature  $T_{re}$ and the  model  parameter $\delta$ in the Jordan frame and the Einstein frame. The value of the reheating temperatures are approximately same. } \label{f2}
\end{figure}

We may compare  this to  the Starobinsky model, $f(R) = R + R^2/6M^2$, where  the  spectral index and tensor-to-scalar ratio in the Jordan frame are \cite{Ia5,Ia9}
\begin{align}
  n_s \simeq 1 -  \dfrac{2}{N_k};
&& r \simeq \dfrac{12}{N_k^2},
\label{IV1.1}\end{align}
with  $ {N}_k \simeq {1}/(2{\varepsilon_1})$,   and  in the Einstein frame
\begin{align}
 \tilde{n_{s}}\simeq 1- \dfrac{2}{\tilde{N}_k}-\dfrac{3}{\tilde{N}_k^2};
&& \tilde{r} \simeq\dfrac{12}{\tilde{N}_k^2},
\label{IV2.1}\end{align}
with  $ \tilde{N}_k = {1}/(2\tilde{\varepsilon_1})$.

The above equations \eqref{IV1.1} and \eqref{IV2.1} clearly show  that the expressions  for spectral index and tensor-to-scalar ratio in the Jordan frame and in the Einstein frame, respectively,   are  not exactly  same. However,  as   the third term containing $3/(\tilde{N}_k^2) $  in  the equation \eqref{IV2.1} is negligible in  comparison to the second term $1/{\tilde{N}_k} $,   equations \eqref{IV1.1} and \eqref{IV2.1} become approximately  equivalent in the Starobinski model.

Similarly,  for our model \eqref{I1}, spectral index and tensor-to-scalar ratio in the Jordan frame and in the Einstein frame are exactly same if $\delta\simeq 1$.
However,  we use   $\delta = 0.98$ from equation \eqref{IIIA10} for calculating  scalar spectral index and tensor-to-scalar ratio in the Jordan frame and in the Einstein frame.   The corresponding  values  turn out, respectively,  as
\begin{align}
 n_s = 0.99972;
&& r = 2.2\times10^{-3},\label{IV3}\\
 \tilde{n}_s = 0.99972
&& \tilde{r} = 2.19\times10^{-3}.
\label{IV4}\end{align}


Thus,  we find  that   the Jordan and the Einstein frame are  equivalent  when  $\delta = 0.98$ as seen  from equations \eqref{IV3} and \eqref{IV4}. A comparison between  both frames is shown   in  Figures 10, 11, 12, 13.     There is an  increasingly   marked departure  from  mutual  equivalence  as  the value of $\delta$  rises  above  $\delta =1$  during the inflation. On the contrary,  from    Fig.10  and  Fig.11,   we note   that when $\delta$  falls  below  $\delta =1$, the differences in the values  of $n_s$ and $r$  in the Jordan frame and  in the Einstein frame  are quite small making the two frames almost indistinguishable from each other.

In Fig.12,  the  maximum and minimum observational values  of $n_s$  at $r = 0.065$ have been shown by the horizontal dashed straight  line.  The solid straight line at the bottom represents   the calculated value of $r \simeq 0.0022$  indicating  $n_s = 0.9997$ in the both frames. The difference between  the   calculated value ($n_s = 0.9997$  at $\delta = 0.98$)  and the observational value ($n_s = 0.9649 \pm 0.0042 $) lies over the range  $\sim 0.0306$  to  $ 0.0390$.

Further, we have the value of  reheating temperature at $\delta = 0.98$ in the Jordan frame and in the Einstein frame,  respectively,  as
\begin{eqnarray} T_{re} = 4.42 M_p \Big(\frac{11g_{s,re}}{43}\Big) \Big(\dfrac{30}{\pi^2 g_{re}} \Big) \Big(\frac{k}{a_o T_o}\Big)^3  \nonumber\\ \Big(\dfrac{\sqrt{2}}{\pi\sqrt{\mathcal{P}_s}r} \Big)e^{3N_k}\dfrac{1}{\sqrt{F}}, \label{IV5}\end{eqnarray}
and
\begin{eqnarray} \tilde{T}_{re} = 4.57 M_p \Big(\frac{11g_{s,re}}{43}\Big) \Big(\dfrac{30}{\pi^2 g_{re}} \Big)\nonumber\\ \Big(\frac{k}{a_o T_o}\Big)^3 \Big(\dfrac{\sqrt{2}}{\pi \sqrt{\tilde{\mathcal{P}}_s}\tilde{r}} \Big) e^{3\tilde{N}_k}. \label{IV6}\end{eqnarray}
 Now,    putting   ${F} = exp(\sqrt{(2/3)}\phi/M_p)$   in  equation  \eqref{IC15} where  $\phi = M_p$ when  the inflation begins and     $\phi \leq M_p$  at the time of horizon exit,    we  obtain   $\tilde{T}_{re}$ in the Einstein frame through
\begin{eqnarray} T_{re} \simeq \frac{1}{\sqrt{F}}\tilde{T}_{re}. \label{IV7}\end{eqnarray}
The  above equation \eqref{IV7} shows the relation between reheating temperature in the Jordan frame and  in the Einstein frame. In both  cases,   $\sqrt{F}\ll 1$ or $\sqrt{F}\gg 1$, these frames do  not remain equivalent.  ${T}_{re}$ and $\tilde{T}_{re}$ strongly  depend  on the number of e-folding  and energy density at the end of the inflation.

Numerical values  of reheating temperature at $\delta = 0.98$ in the Jordan frame and the Einstein frame are given as
\begin{align}
{T}_{re} = 1.98\times10^{17}  GeV;
&& \tilde{T}_{re} = 3.1\times10^{17} GeV,
\label{IV8}\end{align}
and at $\delta \simeq 0.9789$ (or number of e-folding $ N \simeq 57$),
\begin{align}
{T}_{re} = 2.3\times10^{13}  GeV;
&& \tilde{T}_{re} = 3.7\times10^{13} GeV.
\label{IV9}\end{align}
At $\delta \simeq 0.9788$
\begin{align}
{T}_{re} = 5.99\times10^{12}  GeV;
&& \tilde{T}_{re} = 9.3\times10^{12} GeV.
\label{IV10}\end{align}
We can see from the equations \eqref{IV8}, \eqref{IV9} and \eqref{IV10} that even a  small variation  in the value of $\delta $ produces a  drastic change in the reheating temperature $T_{re}$.  However,  both frames provide  approximately   same value of reheating temperature.

Equations  \eqref{IV3}, \eqref{IV4}, \eqref{IV8}, \eqref{IV9} and \eqref{IV10}, are approximately same as a consequence of equivalence  between the Jordan frame and the Einstein frame.  This is also seen from  Fig.13,  where  the  curves  show  the similar  behaviour  leaving only a  small  difference in the value of reheating temperature.

\section{\label{5} Conclusion }
We have ended up this paper   showing  equivalence between   the Jordan frame and the Einstein frame,  while examining the viability of our  model.     We   calculated   the model parameter $\delta$ during the inflation and attempted  to constraint the upper limit on the reheating temperature.  Throughout,  we  have studied the framework of   $f(R) = R^{1+\delta}/R_c^\delta$ model   which provides   the value of the power spectrum indices  very close  to the observations.   Inflationary slow roll parameters, scalar spectral indices and tensor-to-scalar ratio are sensitive to  $\delta$  parameter.   If $\delta= 1$ then $n_s$ becomes  $= 1$  implying  that  tilt of the power spectrum is zero.  However,    considering  that   the Hubble parameter   is  not exactly  constant during inflation,   tilt should be $n_s\sim 1$.  We have calculated value $n_s = 0.99972$ at $\delta = 0.98$ but the observational constraints on the scalar spectral index from   Planck 2018  give  $n_s = 0.9647 \pm 0.0042$. Thus,  there is small difference  between  the observed  and the  calculated value of the $n_s$ in our model, that can be attributed to several factors, including statistical or systematic  errors.  We also see  that tensor-to-scalar ratio  $r$  becomes  zero if $\delta= 1$  which  means   that the   amplitude of the tensor power spectrum is zero. Therefore, primordial  gravitational waves  cannot  not get produced   during inflation if $\delta = 1$,  and    $\delta \simeq 0.98 $  appears consistent with  inflation.

We have the relation \eqref{IIA8} between $n_s$ and $r$  in the Jordan frame in terms of  $\delta$  parameter, while  on the other hand,  the  relation \eqref{IIIA5}  is  completely  independent of $\delta$ in the Einstein frame.  Thus,  it leads to  the same result as equation \eqref{IIA8} at $\delta = 0.98 $.  This can also  be seen from the Fig.12.   Above all, it appears that   $\delta = 0.98 $  is the most preferred value of $\delta$ during the inflation for our  model.

We have obtained another result about $\delta$ parameter which shows   that the  Jordan and the Einstein frame are    equivalent  if $\delta = 0.98$.  In addition,  we found  that  scalar spectral index, tensor-to-scalar ratio and reheating temperature in the Jordan frame show  an increasingly marked difference  from the Einstein frame when  $\delta$ rises above $\delta=1 $. Thus,  a lower  value of this model parameter is favoured.

Calculations show that the reheating temperature at  $\delta = 0.98$  is   $T_{re}\sim 10^{17}$ GeV.  This value of the reheating temperature is required  for grand unification symmetry breaking.  However, it is interesting to see  that  if we  introduce even  a small variation   from $\delta =0.98 $ to $\delta = 0.9788$,   then the reheating temperature hugely drops   to  $T_{re}\sim 10^{13}$ GeV.  Clearly,  this  shows  that the  reheating temperature is  strongly sensitive to  the value of  $\delta$  parameter, although it does not tell us how exactly  reheating occurs.   In our future work,  we will attempt  to calculate reheating temperature via perturbative reheating mechanism or  other  possible  processes   for such  potential.  Further  investigations may be done to study  phase transition and we  guess that some  corrections may   arise  in the potential.   We would  also use this model to examine the evolution of  $\delta$, especially  up to phase of   the late-time cosmic acceleration of the universe.
\begin{center}
\textit{Acknowledgments}
\end{center}
 Authors are thankful to IUCAA, Pune for extending the  facilities  and support  under  the  associateship programme where  most of the work was done.     AS  also thanks  Vipin Sharma, Bal Krishna Yadav, Swagat Mishra and Varun Sahni for the useful   discussions  on  various aspects of  inflationary theories.

 {}
\end{document}